\newcommand{\CCO}{CuCrO$_2$}
\newcommand{\CFO}{CuFeO$_2$}
\newcommand{\MWO}{MnWO$_4$}
\newcommand{\TMO}{TbMnO$_3$}

\newcommand{\NVO}{Ni$_3$V$_2$O$_8$}
\newcommand{\RFMO}{RbFe(MoO$_4$)$_2$}
\newcommand{\LCO}{LiCu$_2$O$_2$}

\def\be{\begin{equation}}
\def\ee{\end{equation}}
\def\bd{\begin{displaymath}}
\def\ed{\end{displaymath}}
\def\Qx{\tau_x}

\documentclass[twocolumn, showpacs, amsmath, amssymb,linenumbers, prb, aps]{revtex4-1}

\usepackage{graphicx}
\usepackage{bm}
\usepackage{ulem}
\usepackage{color}

\begin{document}

\newcommand{\nsub}[1]{_{\rm{#1}}}
\newcommand{\nsup}[1]{^{\rm{#1}}}
\newcommand{\grad}{\ensuremath{^\circ}}
\newcommand{\strike}[1]{\textcolor{blue}{\sout{#1}}}
\newcommand{\comment}[1]{\textcolor{red}{\texttt{#1}}}


\title{Magnetic excitations in the geometric frustrated multiferroic \CCO}

\author{M. Frontzek$^1$}
\altaffiliation{Corresponding author: frontzekmd@ornl.gov}
\author{J. T. Haraldsen$^{2,3,4}$, A. Podlesnyak$^1$,
M. Matsuda$^1$, A. D. Christianson$^1$, R. S. Fishman$^4$, A. S.
Sefat$^4$, Y. Qiu$^{5,6}$, J. R. D. Copley$^5$, S. Barilo$^7$, S. V.
Shiryaev$^7$}
\author{G. Ehlers$^1$}

\affiliation{ $^1$Neutron Scattering Science Division, Oak Ridge
National
Laboratory, Oak Ridge, Tennessee 37831, USA\\
$^2$Theoretical Division, Los Alamos National Laboratory,
Los Alamos, New Mexico 87545, USA\\
$^3$ Center for Integrated Nanotechnologies, Los Alamos National
Laboratory, Los Alamos, New Mexico 87545, USA \\
$^4$Materials Science and Technology Division, Oak Ridge National
Laboratory, Oak Ridge, Tennessee 37831, USA\\
$^5$NIST Center for Neutron Research, National Institute of Standards and Technology, Gaithersburg, Maryland 20899, USA\\
$^6$Department of Materials Science and Engineering, University of
Maryland, College Park, Maryland, 20742, USA \\
$^7$Institute of Solid State and Semiconductor Physics, Minsk 220
072, Belarus }

\date{\today}

\begin{abstract}

In this paper detailed neutron scattering measurements of the
magnetic excitation spectrum of \CCO\ in the ordered state below
$T_{\rm{N1}}=24.2$~K are presented.
The spectra are analyzed using a model Hamiltonian which includes
intralayer-exchange up to the next-next-nearest neighbor and
interlayer-exchange.
We obtain a definite parameter set and show that exchange interaction
terms beyond the next-nearest neighbor are important to describe
the inelastic excitation spectrum.
The magnetic ground state structure generated with our parameter set is in agreement with the structure proposed for \CCO\ from the results
of single crystal diffraction experiments previously published.
We argue that the role of the interlayer exchange is crucial to
understand the incommensurability of the magnetic structure as
well as the spin-charge coupling mechanism.

\end{abstract}


\maketitle

\section{Introduction}
\label{Introduction}

Compounds which exhibit both an ordered magnetic phase and a
ferroelectric phase are termed multiferroics. Especially the
multiferroics where the electric polarization can be controlled with
a magnetic field and vice versa are of continuing interest due to
the potential applications. The most promising candidates for such
controllable multiferroic have been found among the materials with
inherent geometric magnetic frustration.~\cite{Che07}

Different mechanisms leading to spin-charge coupling that have been
discussed in the literature include the magneto-elastic
effect,~\cite{Cha06} the `inverse' Dzyaloshinskii-Moriya
interaction,~\cite{Kat05,Mochi10} and electric dipole induction
through hybridization of $p-d$ orbitals as originally proposed by
Arima.~\cite{Arima07} Spin-charge coupling due to magnetostriction
can occur in collinear commensurate magnetic structures as for
instance observed in $R$Mn$_2$O$_5$, where $R$ is a rare earth
metal.~\cite{Cha06} If magnetic order with non-zero chirality
exists, which may be commensurate or incommensurate with the
lattice, the inverse Dzyaloshinskii-Moriya (DM) interaction induces
(by inversion symmetry breaking) an electric polarization component
perpendicular to the spiral axis and the propagation
vector.~\cite{Kat05}  Systems in which this situation is realized
include \TMO,~\cite{Ken05,Tak08,Wilk09,Ali09,Kaji09,Fab09,Shu10}
\MWO,~\cite{Lau93,Tan08a,Tan09,Sha10} \RFMO,~\cite{Klimin03,Ken07}
\LCO,~\cite{Ma04,Ma05,Xiang07,Hu08,Hu09,Seki08b} and
\NVO.~\cite{Lawes04,Lawes05,Har06} Spin-charge coupling through
Arima's mechanism requires a proper-screw magnetic structure where
the vector of the polarization is parallel to the screw axis and to
the propagation vector, \CFO\ is the most prominent
example.~\cite{Arima07,Uhr96,Ye07,Wang08,Har09a,Nak09,Qui09,Lum10}

In this article, we report a detailed analysis of the spin dynamics
of the multiferroic system \CCO\ which has already been studied
using a variety of techniques such as polarization in applied
magnetic and electric fields,~\cite{Seki08,Kim09} electron spin
resonance (ESR), ~\cite{Yama10} x-ray emission spectroscopy,
(XES)~\cite{Arn09,Shin09} single crystal x-ray
diffraction,~\cite{Kim09b} neutron
diffraction,~\cite{Kad90,Poi09,Soda09,Soda10,Fron11} and inelastic
neutron scattering.~\cite{Kaj10, Poi10} This system is isostructural
to \CFO\ and a detailed comparison of the two systems is
instructive.

In contrast to \CFO\ which becomes multiferroic in an applied
magnetic field ~\cite{Kim06} or through doping the Fe-site with
Al,~\cite{Seki07} Ga~\cite{Har10} or Rh~\cite{Kun09}, \CCO\ enters
the multiferroic state in zero field with the magnetic transition.
In both compounds the magnetic structure in the multiferroic phase
is an incommensurate proper-screw magnetic structure. However, the
propagation vector found for \CCO\ with $\bm{\tau}=(\tau,\tau,0)$
and $\tau=0.3298(1)$ is very close to the commensurate value. Unlike
the propagation vector of \CFO\, which in comparison is very
different, $\bm{\tau}=(\tau,\tau,3/2)$ with
$\tau=0.207$.~\cite{Nak07}

\section{Experimental}
\label{Experimental}

A detailed account of the sample preparation was given
previously.~\cite{Fron11} The trigonal crystal structure (space
group $R\bar{3}m$) with lattice parameters $a=2.97$~{\AA} and
$c=17.110$~{\AA} was confirmed by x-ray powder analysis of crushed
crystals. Further characterization with respect to their magnetic
properties was done using a SQUID-magnetometer. The obtained
susceptibility curves are similar to data published
previously.~\cite{Kim08,Poi09,Kim09,Soda10} Identifying the same
characteristic points in the susceptibility data as Kimura et
al.~\cite{Kim08} the same two characteristic phase transition
temperatures, $T\nsub{N1}=24.2$~K and $T\nsub{N2}=23.6$~K, were
obtained for our samples. The Curie-Weiss fit between 148~K and
287~K of the inverse susceptibility gave an asymptotic paramagnetic
Curie temperature of -200(1)~K and an effective moment of
3.88(1)~$\mu\nsub{B}$ per Cr$^{3+}$ ion. Measurements of the
magnetization measured along three orthogonal directions, [$110$],
[$\overline{1}10$] and [$001$], are shown in Fig.~\ref{SuscFig}
below. A phase transition at $H\nsub{flop}\sim5.3$~T can be seen in
these data (the value is determined from the center of gravity of
the peak in the derivative), in agreement with earlier
reports.~\cite{Kim09} At this phase transition the electrical
polarization is flopped~\cite{Kim09} in conjunction with a
reorientation of the ordered magnetic moments.~\cite{Soda10}

\begin{figure}[h]
\includegraphics[width=8cm]{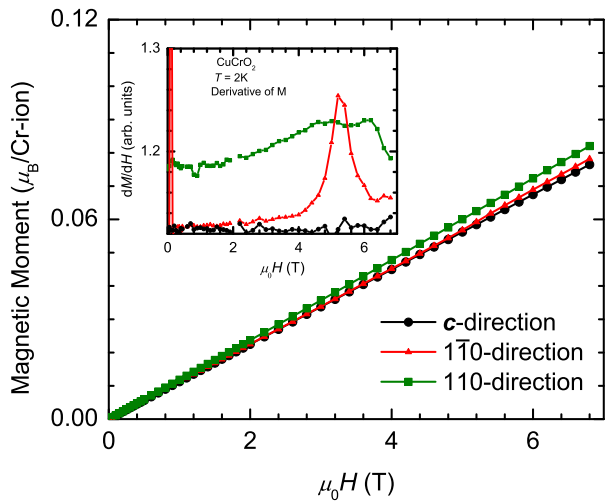}
\caption{(Color online) Magnetization measurement along the three
main crystallographic directions in \CCO\ single crystals at
$T$~=~2~K. The inset shows the derivative of the magnetization with
a peak at $H\nsub{flop}$ in the [$\overline{1}10$] direction.}
\label{SuscFig}
\end{figure}

Ten crystals with a total mass of $m\sim{0.6}$~g were co-aligned on
an aluminum sheet covering an area of approx. $20\times{20}$~mm for
inelastic neutron scattering experiments. The crystals were platelet
like with the $c$-direction normal to the plate surface. The
horizontal scattering plane was $HHL$. Experiments were conducted at
the Cold Neutron Chopper Spectrometer (CNCS) at the Spallation
Neutron Source in Oak Ridge~\cite{Mason06}, the HB-1 triple-axis
spectrometer at the High Flux Isotope Reactor in Oak Ridge, and at
the Disk Chopper Spectrometer (DCS) at the NIST Center for Neutron
Research (NCNR).~\cite{Copley03}

All experiments used a standard orange cryostat in a temperature
range from 1.5 to $\sim{100}$~K. The CNCS measurements were
performed in two settings with different incident neutron energies,
12.1 meV and 3 meV, respectively. The energy resolution at the
elastic line was 0.4350(6)~meV full width at half max. (FWHM) at
12.1 meV and 0.0649(1)~meV FWHM at 3 meV, respectively. The HB-1
measurements used constant $k_{f}=14.7$ meV which resulted in an
effective energy resolution of 1.84 meV at 7.5 meV. The collimation
was 48-60-60-240 with two additional pyrolitic graphite (PG) filters
to suppress higher order contamination. The DCS measurement was
performed with an incident energy of 3.53 meV with a measured
resolution of 0.1 meV (FWHM) at the elastic line. The data obtained
on CNCS and DCS have been reduced using the DAVE software
package.~\cite{Dave}

\section{Theory}
\label{Theory}

\begin{figure}[b]
\includegraphics[height=7cm]{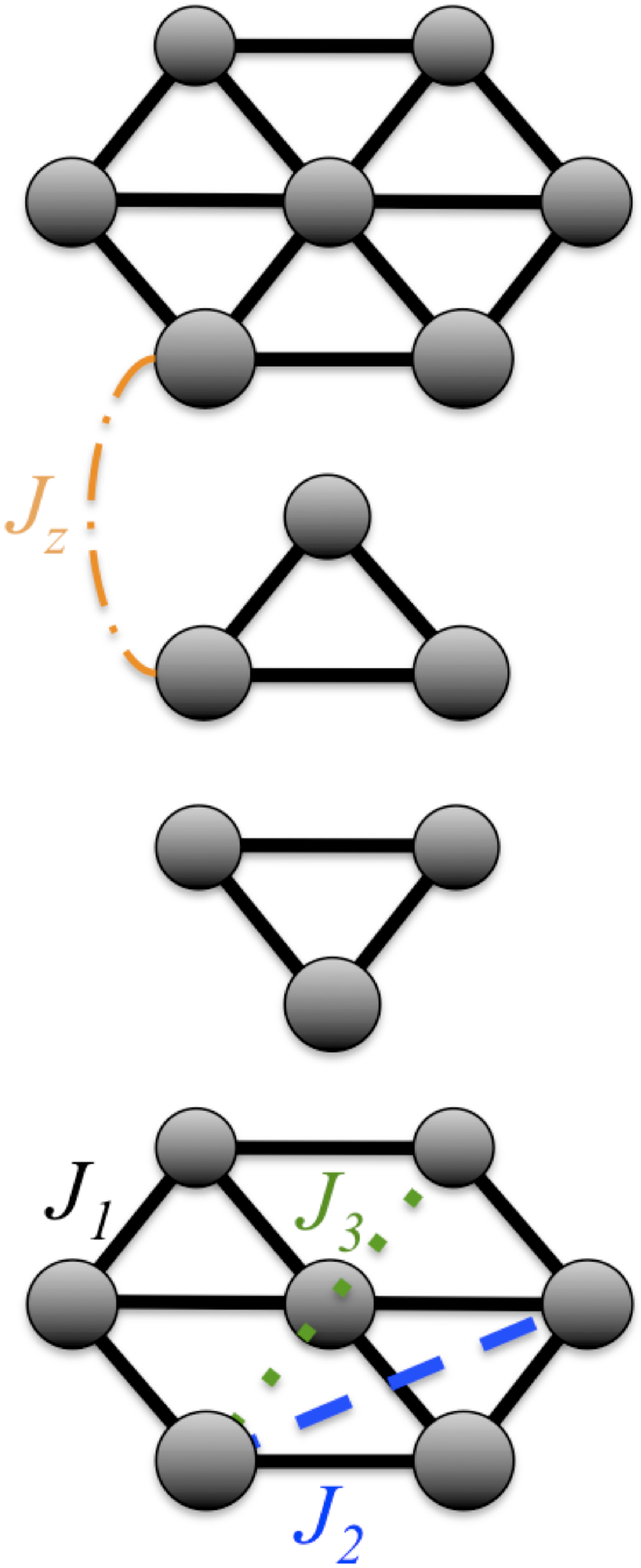}
\caption{(Color online) Considered exchange paths in the Heisenberg
Hamiltonian.}  \label{paths}
\end{figure}

The hexagonal symmetry of the \CCO\ lattice provides a complex
network of possible intra- and inter-layer superexchange
pathways~\cite{Fish08} that are described by the Heisenberg
Hamiltonian \be H = -\frac{1}{2}\sum_{i \neq j} J_{ij}
\mathbf{{S}}_i \cdot \mathbf{{S}}_j - D_x \sum_i \mathbf{{S}}_{ix}^2
- D_z \sum_i \mathbf{{S}}_{iz}^2{\;}, \label{genH} \ee where
$\mathbf{S}_i$ is the local moment on site $i$. The superexchange
interactions $J_{ij}$ between sites $i$ and $j$ are
antiferromagnetic when $J_{ij}<0$. An overview of the exchange paths
in respect to the lattice is given in Fig.~\ref{paths}. The
single-ion anisotropy along the $x$ and $z$ axes is given by
$D_{x,z}$, where $D>0$ produces easy-axis anisotropy and $D<0$
produces easy-plane anisotropy, respectively. The three-dimensional
magnetic state is constructed by stacking the two-dimensional
configurations ferromagnetically along the $c$-axis.

Through an energy minimization of the exchange parameters and
anisotropy, the magnetic ground state configuration is determined
through a classical approach described in Ref.~\onlinecite{Fish10}
by defining $S_z$ within any hexagonal plane as \be \displaystyle
S_z(\mathbf{R}) = A \cdot \sum_{l=0}C_{2l+1}\cos[\Qx(2l+1)x] \ee
where the $C_{2l+1}$ harmonics are produced by the easy axis
anisotropy $D_z$. With $C_1$ set to 1, the amplitude $A$ is obtained
from the condition that the maximum value of $|S_z(\mathbf{R})|$
equals $S$. The perpendicular spin components $S_y$ are given by \be
\displaystyle S_y(\mathbf{R}) = \sqrt{S-S_z(\mathbf{R})^2} \, \cdot
\, {\rm{sgn}} (\sin(\Qx x)) {\;}. \ee The ordering wavevector $\Qx$
and coefficients $C_{2l+1}$ are determined by minimizing the energy
on a large unit cell of size
$\sim{10}^{4}{\;}{a}\times{a}\times{c}$, where $a$ is the lattice
constant within a hexagonal plane and $c$ is the separation between
neighboring planes.

Based on this magnetic ground state, the spin dynamics are evaluated
using a Holstein-Primakoff transformation, where the spin operators
are given by $S_{iz} = S -a_{i}^{\dag} a_{i}$, $S_{i+} =
\sqrt{2S}a_{i}$, and $S_{i-} = \sqrt{2S}a_{i}^{\dag}$ ($a_i$ and
$a_i^{\dag }$ are boson destruction and creation operators). A
rotation of the local spin operators accounts for the
non-collinearity of the spins.~\cite{Har09b,Zhito96}

To determine the spin wave (SW) frequencies $\omega_{\mathbf{Q}}$,
we solve the equation-of-motion for the vectors $\mathbf{v_{Q}} =
[a_{\mathbf{Q}}^{(1)},
a_{\mathbf{Q}}^{(1)\dag},a_{\mathbf{Q}}^{(2)},
a_{\mathbf{Q}}^{(2)\dag},...]$, which may be written in terms of the
$2N \times 2N$ matrix $ \underline{M}(\mathbf{Q})$ as
$id\mathbf{v_Q}/dt = -\big[ \underline{H}_{2},\mathbf{v_Q}\big] =
\underline{M}(\mathbf{Q}) \mathbf{v_Q}$, where $N$ is the number of
spin sites in the unit cell.~\cite{Har09b} The SW frequencies are
then determined from the condition Det[$ \underline{M}(\mathbf{Q}) -
\omega_{\mathbf{Q}} \underline{I}$] = 0. To assure the local
stability of a magnetic phase, all SW frequencies must be real and
positive and all SW weights must be positive.

The SW intensities or weights are coefficients of the spin-spin
correlation function: \be
\begin{array}{c}
\displaystyle S(\mathbf{Q},\omega) = \sum_{\alpha \beta}
(\delta_{\alpha \beta} - Q_{\alpha}Q_{\beta})S^{\alpha
\beta}(\mathbf{Q},\omega),
\end{array}
\label{weights} \ee where $\alpha$ and $\beta$ are $x$, $y$, or
$z$.~\cite{Zhito96} A more detailed discussion of this method is
contained in Ref.~\onlinecite{Har09b}. Notice that magnetic neutron
scattering measurements (INS) only detect components of the spin
fluctuations perpendicular to the wavevector $\mathbf{Q}$. The total
intensity $I(\mathbf{Q},\omega)$ for an INS scan at constant
$\mathbf{Q}$ is given by \be I(\mathbf{Q},\omega) =
S(\mathbf{Q},\omega) F_{\mathbf{Q}}^2\exp
\bigl(-(\omega-\omega_\mathbf{Q})^2/2\delta^2 \bigr) {\;}, \ee where
$\delta$ is the energy resolution and $F_{{\bf Q}}$ is the Cr$^{3+}$
magnetic form factor.

This approach yields additional information on the magnetic ground
state. The magnetic ground state is not provided for these systems
and must therefore be derived from the energy minimization of the
Hamiltonian possible magnetic structures within the
$\sim{10}^{4}{\;}{a}\times{a}\times{c}$ cell. Therefore, two
energetically degenerate states, for instance commensurate vs.
slightly incommensurate, can be distinguished.

\section{Results}
\label{Results}

\begin{figure}[h]
\includegraphics[width=9cm]{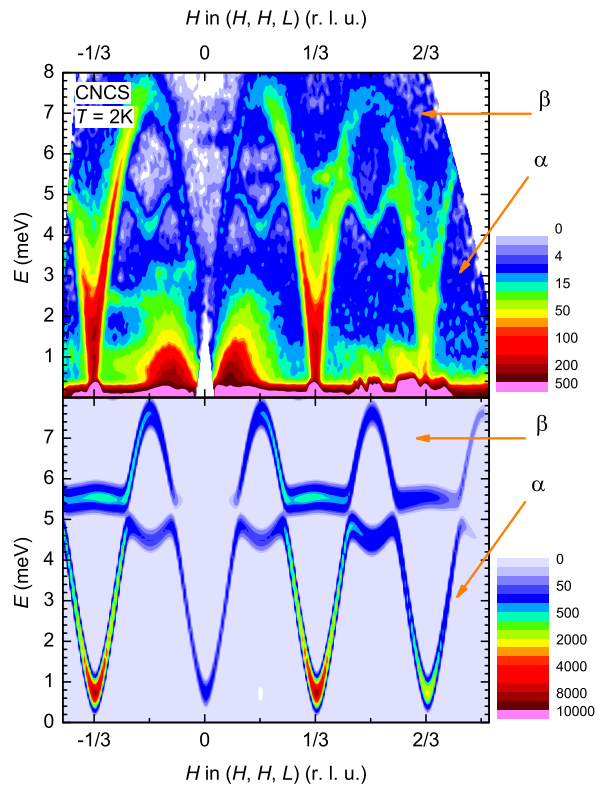}
\caption{(Color online) Upper panel: Magnetic excitation spectrum in
$S(\mathbf{Q},\omega)$ of CuCrO$_2$ measured at $T=2$~K at CNCS.
Integration range along $L$ was from 0 to 5 in r.~l.~u., and along
the $H\overline{H}$ direction $\pm{0.025}$ r.~l.~u.. The intensity
around $H=0$ at low energy originates from the halo of the primary
beam. Lower panel: Spin waves computed from the best theoretical
model, the modes discussed in the text are marked $\alpha$,
$\beta$.} \label{Map12meV}
\end{figure}

The inelastic excitation spectrum of \CCO\ in the $HH$ direction as
measured at CNCS with $E\nsub{i}=12$~meV is shown in the upper panel
of Fig.~\ref{Map12meV}. Integration along the $L$ direction
was in the range $0<L<5$ r. l. u. (relative lattice units) which is
justified by a rather small dispersion along this direction.
Integration along the perpendicular $H\overline{H}$ direction was
within $\pm{0.025}$ r. l. u. (corresponding to $\pm{2.5}$ deg. out
of the scattering plane). For comparison the model calculation is shown
in the lower panel.

The low energy mode $\alpha$ originates from the magnetic Bragg peak
in the vicinity of $H=1/3$ and flattens off at around 5 meV. It has
a cusp like local energy minimum at the magnetic zone boundary at
$H=1/6$. The intensity of this mode is strongest in the vicinity of
the Bragg peak and falls off towards the zone boundary. This mode is
mainly influenced by the model parameters $J_2$, $J_3$, $D_x$ and
$D_z$ (see above). The minimum of the $\alpha$ mode at $H=1/6$ is of
considerable interest. It can only be modeled with the inclusion of
an antiferromagnetic next-next nearest neighbor exchange interaction
$J_3$. If $J_3$ is neglected or ferromagnetic, the excitation would
be flat at $H=1/6$ or would show a local maximum. Analyzing the
intensity of the $\alpha$ mode at the zone boundary, the measurement
shows more intensity at $H=1/2$ than at $H=1/6$. In the modeling
this leads to a negative in-plane anisotropy constant $D_x$
(otherwise the intensity would be higher at $H=1/6$). In return,
this leads to a ground state with a proper screw magnetic structure
rather than a cycloid.

\begin{figure}[h]
\includegraphics[width=8cm]{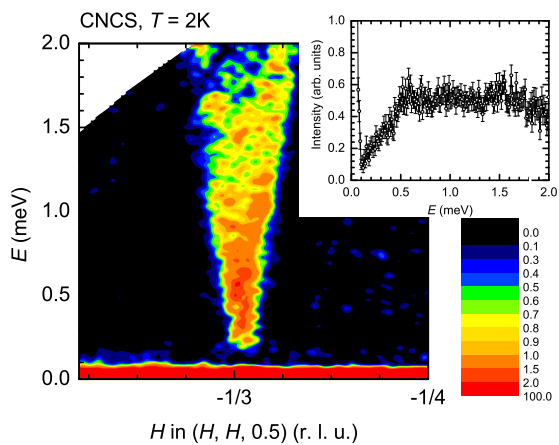}
\caption{(Color online) Magnetic excitation spectrum of \CCO\
measured at $T=2$~K at CNCS with 3 meV incident energy. The inset
shows a constant-$Q$ cut along the excitation. Error bars represent
$\pm{1}\sigma$ from counting statistics.} \label{Map3meV}
\end{figure}

The non-zero anisotropy terms $D_x$ and $D_z$ mean that the $\alpha$
mode must be gapped. The gap is too small to be unambiguously
detected at $E\nsub{i}=12.1$ meV. However, with improved energy
resolution ($E\nsub{i}=3$ meV) a gap of $\sim{0.5}$~meV is clearly
seen as shown in Fig.~\ref{Map3meV}. Here the integration along the
$L$-direction is only for a small range around $L=1$. The absolute
values of $D_x$ and $D_z$ are adapted in the theoretical
calculations to accurately model this gap.

An overall weaker and flat $\beta$ mode is observed between 5 and 8
meV. The measurement did not resolve whether a crossing of the
$\alpha$ and $\beta$ mode occurs as suggested by the calculation,
mainly due to insufficient resolution. The $\beta$ mode has a
maximum of $\sim{7.5}$~meV at the magnetic zone boundaries at
$H=1/6$ and $H=1/2$. The energy of the $\beta$ mode at these points
is mainly determined by $J_2$ and to a lesser degree by $J_3$.
Kajimoto et al.~\cite{Kaj10} ascribed the $\beta$ mode (referred to
as ``flat component'') to the existence of an interlayer exchange
interaction $J_z$ which is inconsistent with our data. In the lower
panel of Fig. 3, the computed spin wave excitation spectrum form the
best theoretical model is shown. The $\alpha$ and $\beta$ mode in
this energy range determine $J_2$ and $J_3$ as well as $J_1$ to
which all parameters are relative. In agreement with data from the
literature,~\cite{Kaj10, Poi10} a survival of magnetic collective
dynamics up to several times $T_{\rm{N}}$ is observed at the
position of the $\alpha$ mode.

\begin{figure}[h]
\includegraphics[width=8.5cm]{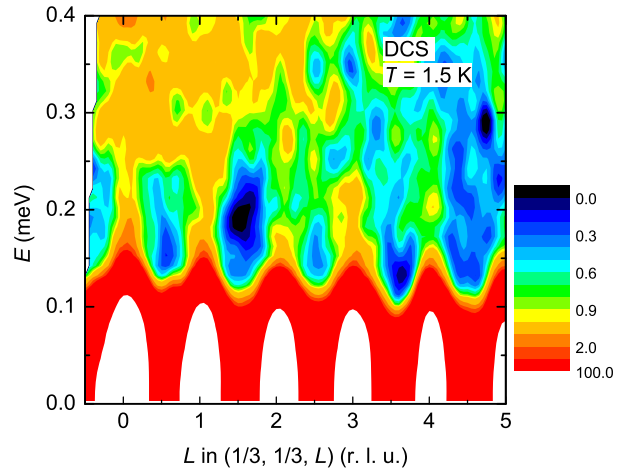}\\
\caption{Magnetic excitation spectrum in  $S(\boldmath{Q}, \omega)$
of CuCrO$_2$ measured at $T$ = 2 K at DCS with 3.55 meV incident
energy. The data is integrated in the $HH$ of 0.32 to 0.34 r. l. u.
from the central detector bank. The intensity is color coded in
a linear scale with the exception of the elastic Bragg peaks with
two orders of magnitude higher intensity.}\label{DCS}
\end{figure}

The spin-wave spectrum along the $L$-direction is dispersion-less
for energies above 0.5 meV as already mentioned above. However,
below the energy gap of 0.5 meV a modulation can be seen
Fig.~\ref{DCS}. For an energy transfer of 0.2 meV, the measured
intensity along $L$ is higher at the position of the magnetic Bragg
peaks compared to the position between.
This intensity pattern can be reproduced with the introduction of a
ferromagnetic interlayer coupling $J_z$. The magnitude of the
interlayer exchange is small as is the effect on the excitation
spectrum.

The data presented so far allow the determination of the values for
the exchange interaction and the anisotropy terms within the given
model. The calculations replicate satisfactorily the $\alpha$ and
$\beta$ excitation modes as shown in the lower panel of
~\ref{Map12meV}. The intensity pattern of the DCS measurement
(Fig.~\ref{DCS}) is modeled with the small interaction term $J_z$.
The interlayer exchange $J_z$ also results in the magnetic ground
state with the incommensurate ordering wavevector $\Qx=0.329$.
Without the interlayer exchange the magnetic ground state would be
commensurate. The model Hamiltonian also reproduces the gap in the
excitation spectrum, using the anisotropy terms, which as a
consequence leads to the splitting of the otherwise degenerated
magnetic ground state. This splitting of the degenerate ground state
gives rise to another excited state $\beta$' at higher energies,
with a spin wave dispersion that mirrors the $\beta$ mode from the
ground state but which has an additional gap of 2.2~meV. The
intensity of this mode is weaker than the excitations from the
ground state and cannot be seen in the CNCS data, likely because, by
way of how the $(\mathbf{Q},\omega)$ space is mapped in a
time-of-flight measurement with the chosen settings, only $L>1$ is
covered at $\hbar\omega\gtrsim{8}$~meV.

\begin{figure}[h]
\includegraphics[width=8cm]{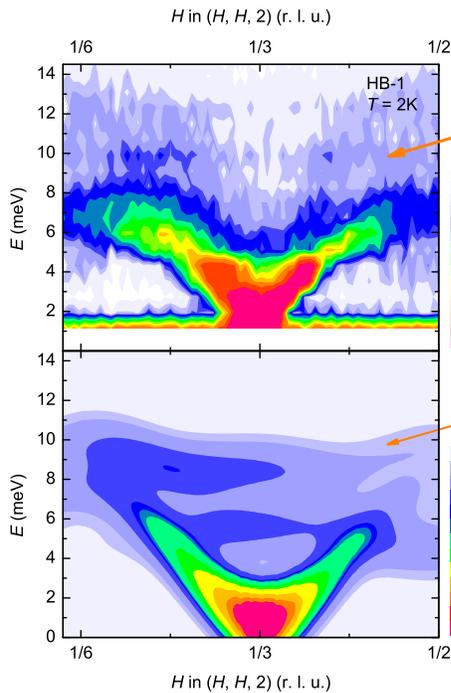}
\caption{(Color online) Upper panel: Contour map from constant-$E$
scans of \CCO\ measured at $T=2$~K at HB-1. Lower panel: The
corresponding model of the $\alpha$, $\beta$ and $\beta$'
excitations.} \label{HB1}
\end{figure}

Figure ~\ref{HB1} shows a contour map of the measurements taken at
HB-1. These are constant-$E$ scans with an energy difference of 0.5
meV in the range from 1.5~meV to 15~meV. The measurements are along
the ($HH2$) direction. In this figure, it can be seen that another
mode with nearly the same dispersion exists above the $\beta$ mode,
which we identify with the $\beta$' mode resulting from the
calculations. The coarser energy resolution of HB-1 leads to a
partial blur of the $\beta$ and $\beta$' mode. The calculation
yields a gap between both modes of 2.2~meV at the zone boundary.

To summarize the results, the intensity and dispersion of
experimentally observed spin-wave modes in \CCO\ have been modeled
with a Hamiltonian that includes at least six free parameters, which
are given in Table~\ref{Table1}.
\begin{table}[h]
\begin{center}
\renewcommand{\arraystretch}{1.5}
\renewcommand{\tabcolsep}{0.2cm}
\begin{tabular}{|c|c|c|c|c|c|c|}
\hline Set & $J_1$ & $J_2$  & $J_3$ & $J_z$ &
$D_x$  & $D_z$  \\
\hline Ref.~\onlinecite{Poi10} & -2.3 & -0.12 & - & - & -0.4$^*$ & 0.4$^*$ \\
\hline This work & -2.8 & -0.48 & -0.08 & 0.02 & -0.59 & 0.48 \\
\hline CuFeO$_2$ & -0.23 & -0.12 & -0.16 & -0.06$^{\dag}$ & - & 0.22 \\
 \hline
\end{tabular}
\caption{Comparison of the relevant exchange interaction and
anisotropy parameters from Ref.~\onlinecite{Poi10} ($^*$only one
value was fitted) with this work and the results for \CFO\ from
Ref.~\onlinecite{Fish08}($^{\dag}J_{z1}$). Energies are in meV.
\label{Table1}}
\end{center}
\end{table}

Small discrepancies between calculation and measurement suggest the
need to include higher order parameters beyond the ones used here.
This is most apparent in the slight discrepancy of the spin-wave
velocities. The velocities depend in a non-trivial way from all
interactions and deviations from the model may indicate the need for
magneto-elastic or bi-quadratic terms. While the addition of $J_3$
and $D_z$ helps reduce this difference, it is clear that other
interactions may be affecting the system. The deduction of the
parameters in the Hamiltonian has been based on the approach to
incorporate the least necessary number to describe the excitation
spectrum satisfactorily.

In comparison to CuFeO$_2$, the nearest neighbor intralayer exchange
interaction $J_1$ is one order of magnitude stronger in CuCrO$_2$,
but the interlayer exchange and the anisotropy parameter $D_z$ are
of comparable magnitude.~\cite{Har10b} The different magnetic ground
states are explainable with the different ratio of $D/|J_1|$. In
CuCrO$_2$, where this ratio is small, the proper-screw is the stable
magnetic structure, while in \CFO\ the four-sublattice collinear
structure is the ground state.~\cite{Fish10} It has been interpreted
that the main effect of doping in \CFO\ is the decrease of
anisotropy and through this the proper-screw magnetic structure can
be stabilized as ground state in the doped compounds.~\cite{Har10}
Notably is the difference of the in-plane anisotropy $D_x$ which is
absent in \CFO\ where a Goldstone mode at the incommensurate
wavevector is observed~\cite{Fish08}, but present in \CCO\ as
indicated by the gap of the $\alpha$ mode. Instead of $D_x$ the
observed lattice distortion in the basal plane is relevant to model
the excitation spectra in \CFO.~\cite{Har10b}

The interlayer exchange in \CFO\ leads to a 10-sub lattice stacking
sequence along the $c$-direction and can be modeled with one
ferromagnetic and two antiferromagnetic exchange
parameters.~\cite{Fish08} The interlayer exchange in \CCO\ seems
simpler and can be described with one ferromagnetic parameter of
similar magnitude. In \CFO\ the interlayer exchange has been the
most affected parameter by doping~\cite{Har10b} which might explain
the difference between \CCO\ and CuFeO$_2$.

The last marked difference to be discussed is the apparent absence
of a structural phase transition in CuCrO$_2$. Strain measurements
on \CCO\ ~\cite{Kim09b} indicate strong magnetoelastic coupling, but
apparently insufficient to lead to a phase transition as in
CuFeO$_2$. In the latter, it has been demonstrated that the
inclusion of bi-quadratic terms in the Hamiltonian are relevant in
the prediction of the phase diagram.~\cite{Plu07} In CuCrO$_2$, the
bi-quadratic terms seem less relevant for the understanding of the
magnetic ground state but probably cause the slight discrepancy of
the spin-wave velocities between model and experiment.

\section{Conclusion}
\label{Conclusion}

A detailed investigation of the magnetic excitation spectrum of
CuCrO$_2$, at low temperatures has been performed using neutron
scattering techniques. The excitation spectrum has been used to
deduce the relevant exchange interaction and anisotropy parameters.
The parameter set points to a ground state with an incommensurate
proper-screw magnetic structure in agreement with results published
earlier.~\cite{Poi09,Poi10,Fron11} Antiferromagnetic intralayer
exchange has to be considered up to next-next nearest neighbor in
order to be consistent with the experimental data.

We have also shown that interlayer exchange is relevant for \CCO\
which can thus no longer be considered as a quasi two-dimensional
system. The multiferroic properties of \CCO\ have been explained
within the light of the Arima model which does not consider order
between the spiral planes. It is an interesting question in which
way the interlayer exchange interaction in \CCO\ affects its
multiferroic properties.

\begin{acknowledgments}
We acknowledge the technical and scientific support from the staff
at SNS, HFIR, and NIST. This research was sponsored by the Division
of Materials Sciences and Engineering of the U. S. Department of
Energy. This work utilized facilities supported in part by the
National Science Foundation under Agreement No. DMR-0944772.
Research at Oak Ridge National Laboratory's Spallation Neutron
Source was supported by the Scientific User Facilities Division,
Office of Basic Energy Sciences, U. S. Department of Energy. Some
theoretical aspects of this work has been supported by the Center
for Integrated Nanotechnologies, a U.S. Department of Energy, Office
of Basic Energy Sciences user facility. Los Alamos National
Laboratory is operated by Los Alamos National Security, LLC, for the
National Nuclear Security Administration of the U.S. Department of
Energy under contract DE-AC52-06NA25396. The work in Minsk was
supported in part by Belarusian Fund for Basic Scientific Research,
grant No F10R-154.

\end{acknowledgments}

\pagebreak


\begin{thebibliography}{62}%
\makeatletter
\providecommand \@ifxundefined [1]{%
 \@ifx{#1\undefined}
}%
\providecommand \@ifnum [1]{%
 \ifnum #1\expandafter \@firstoftwo
 \else \expandafter \@secondoftwo
 \fi
}%
\providecommand \@ifx [1]{%
 \ifx #1\expandafter \@firstoftwo
 \else \expandafter \@secondoftwo
 \fi
}%
\providecommand \natexlab [1]{#1}%
\providecommand \enquote  [1]{``#1''}%
\providecommand \bibnamefont  [1]{#1}%
\providecommand \bibfnamefont [1]{#1}%
\providecommand \citenamefont [1]{#1}%
\providecommand \href@noop [0]{\@secondoftwo}%
\providecommand \href [0]{\begingroup \@sanitize@url \@href}%
\providecommand \@href[1]{\@@startlink{#1}\@@href}%
\providecommand \@@href[1]{\endgroup#1\@@endlink}%
\providecommand \@sanitize@url [0]{\catcode `\\12\catcode
`\$12\catcode
  `\&12\catcode `\#12\catcode `\^12\catcode `\_12\catcode `\%12\relax}%
\providecommand \@@startlink[1]{}%
\providecommand \@@endlink[0]{}%
\providecommand \url  [0]{\begingroup\@sanitize@url \@url }%
\providecommand \@url [1]{\endgroup\@href {#1}{\urlprefix }}%
\providecommand \urlprefix  [0]{URL }%
\providecommand \Eprint [0]{\href }%
\@ifxundefined \urlstyle {%
  \providecommand \doi  [0]{\begingroup \@sanitize@url \@doi}%
  \providecommand \@doi [1]{\endgroup \@@startlink {\doibase
  #1}doi:\discretionary {}{}{}#1\@@endlink }%
}{%
  \providecommand \doi  [0]{doi:\discretionary{}{}{}\begingroup
  \urlstyle{rm}\Url }%
}%
\providecommand \doibase [0]{http://dx.doi.org/}%
\providecommand \Doi [0]{\begingroup \@sanitize@url \@Doi }%
\providecommand \@Doi  [1]{\endgroup\@@startlink{\doibase#1}\@@Doi}%
\providecommand \@@Doi [1]{#1\@@endlink}%
\providecommand \selectlanguage [0]{\@gobble}%
\providecommand \bibinfo  [0]{\@secondoftwo}%
\providecommand \bibfield  [0]{\@secondoftwo}%
\providecommand \translation [1]{[#1]}%
\providecommand \BibitemOpen [0]{}%
\providecommand \bibitemStop [0]{}%
\providecommand \bibitemNoStop [0]{.\EOS\space}%
\providecommand \EOS [0]{\spacefactor3000\relax}%
\providecommand \BibitemShut  [1]{\csname bibitem#1\endcsname}%
\bibitem [{\citenamefont {Cheong}\ and\ \citenamefont
  {Mostovoy}(2007)}]{Che07}%
  \BibitemOpen
  \bibfield  {author} {\bibinfo {author} {\bibfnamefont {S.-W.}\ \bibnamefont
  {Cheong}}\ and\ \bibinfo {author} {\bibfnamefont {M.}~\bibnamefont
  {Mostovoy}},\ }\href@noop {} {\bibfield  {journal} {\bibinfo  {journal}
  {Nature Materials},\ }\textbf {\bibinfo {volume} {6}},\ \bibinfo {pages} {13}
  (\bibinfo {year} {2007})}\BibitemShut {NoStop}%
\bibitem [{\citenamefont {Chapon}\ \emph {et~al.}(2006)\citenamefont {Chapon},
  \citenamefont {Radaelli}, \citenamefont {Blake}, \citenamefont {Park},\ and\
  \citenamefont {Cheong}}]{Cha06}%
  \BibitemOpen
  \bibfield  {author} {\bibinfo {author} {\bibfnamefont {L.~C.}\ \bibnamefont
  {Chapon}}, \bibinfo {author} {\bibfnamefont {P.~G.}\ \bibnamefont
  {Radaelli}}, \bibinfo {author} {\bibfnamefont {G.~R.}\ \bibnamefont {Blake}},
  \bibinfo {author} {\bibfnamefont {S.}~\bibnamefont {Park}}, \ and\ \bibinfo
  {author} {\bibfnamefont {S.-W.}\ \bibnamefont {Cheong}},\ }\href@noop {}
  {\bibfield  {journal} {\bibinfo  {journal} {Phys. Rev. Lett.},\ }\textbf
  {\bibinfo {volume} {96}},\ \bibinfo {pages} {097601} (\bibinfo {year}
  {2006})}\BibitemShut {NoStop}%
\bibitem [{\citenamefont {Katsura}\ \emph {et~al.}(2005)\citenamefont
  {Katsura}, \citenamefont {Nagaosa},\ and\ \citenamefont {Balatsky}}]{Kat05}%
  \BibitemOpen
  \bibfield  {author} {\bibinfo {author} {\bibfnamefont {H.}~\bibnamefont
  {Katsura}}, \bibinfo {author} {\bibfnamefont {N.}~\bibnamefont {Nagaosa}}, \
  and\ \bibinfo {author} {\bibfnamefont {A.~V.}\ \bibnamefont {Balatsky}},\
  }\href@noop {} {\bibfield  {journal} {\bibinfo  {journal} {Phys. Rev.
  Lett.},\ }\textbf {\bibinfo {volume} {95}},\ \bibinfo {pages} {057205}
  (\bibinfo {year} {2005})}\BibitemShut {NoStop}%
\bibitem [{\citenamefont {Mochizuki}\ and\ \citenamefont
  {Furukawa}(2010)}]{Mochi10}%
  \BibitemOpen
  \bibfield  {author} {\bibinfo {author} {\bibfnamefont {M.}~\bibnamefont
  {Mochizuki}}\ and\ \bibinfo {author} {\bibfnamefont {N.}~\bibnamefont
  {Furukawa}},\ }\href@noop {} {\bibfield  {journal} {\bibinfo  {journal}
  {Phys. Rev. Lett.},\ }\textbf {\bibinfo {volume} {105}},\ \bibinfo {pages}
  {187601} (\bibinfo {year} {2010})}\BibitemShut {NoStop}%
\bibitem [{\citenamefont {Arima}(2007)}]{Arima07}%
  \BibitemOpen
  \bibfield  {author} {\bibinfo {author} {\bibfnamefont {T.}~\bibnamefont
  {Arima}},\ }\href@noop {} {\bibfield  {journal} {\bibinfo  {journal} {Journal
  of the Physical Society of Japan},\ }\textbf {\bibinfo {volume} {76}},\
  \bibinfo {pages} {073702} (\bibinfo {year} {2007})}\BibitemShut {NoStop}%
\bibitem [{\citenamefont {Kenzelmann}\ \emph {et~al.}(2005)\citenamefont
  {Kenzelmann}, \citenamefont {Harris}, \citenamefont {Jonas}, \citenamefont
  {Broholm}, \citenamefont {Schefer}, \citenamefont {Kim}, \citenamefont
  {Zhang}, \citenamefont {Cheong}, \citenamefont {Vajk},\ and\ \citenamefont
  {Lynn}}]{Ken05}%
  \BibitemOpen
  \bibfield  {author} {\bibinfo {author} {\bibfnamefont {M.}~\bibnamefont
  {Kenzelmann}}, \bibinfo {author} {\bibfnamefont {A.~B.}\ \bibnamefont
  {Harris}}, \bibinfo {author} {\bibfnamefont {S.}~\bibnamefont {Jonas}},
  \bibinfo {author} {\bibfnamefont {C.}~\bibnamefont {Broholm}}, \bibinfo
  {author} {\bibfnamefont {J.}~\bibnamefont {Schefer}}, \bibinfo {author}
  {\bibfnamefont {S.~B.}\ \bibnamefont {Kim}}, \bibinfo {author} {\bibfnamefont
  {C.~L.}\ \bibnamefont {Zhang}}, \bibinfo {author} {\bibfnamefont {S.-W.}\
  \bibnamefont {Cheong}}, \bibinfo {author} {\bibfnamefont {O.~P.}\
  \bibnamefont {Vajk}}, \ and\ \bibinfo {author} {\bibfnamefont {J.~W.}\
  \bibnamefont {Lynn}},\ }\href@noop {} {\bibfield  {journal} {\bibinfo
  {journal} {Phys. Rev. Lett.},\ }\textbf {\bibinfo {volume} {95}},\ \bibinfo
  {pages} {087206} (\bibinfo {year} {2005})}\BibitemShut {NoStop}%
\bibitem [{\citenamefont {Takahashi}\ \emph {et~al.}(2008)\citenamefont
  {Takahashi}, \citenamefont {Kida}, \citenamefont {Yamasaki}, \citenamefont
  {Fujioka}, \citenamefont {Arima}, \citenamefont {Shimano}, \citenamefont
  {Miyahara}, \citenamefont {Mochizuki}, \citenamefont {Furukawa},\ and\
  \citenamefont {Tokura}}]{Tak08}%
  \BibitemOpen
  \bibfield  {author} {\bibinfo {author} {\bibfnamefont {Y.}~\bibnamefont
  {Takahashi}}, \bibinfo {author} {\bibfnamefont {N.}~\bibnamefont {Kida}},
  \bibinfo {author} {\bibfnamefont {Y.}~\bibnamefont {Yamasaki}}, \bibinfo
  {author} {\bibfnamefont {J.}~\bibnamefont {Fujioka}}, \bibinfo {author}
  {\bibfnamefont {T.}~\bibnamefont {Arima}}, \bibinfo {author} {\bibfnamefont
  {R.}~\bibnamefont {Shimano}}, \bibinfo {author} {\bibfnamefont
  {S.}~\bibnamefont {Miyahara}}, \bibinfo {author} {\bibfnamefont
  {M.}~\bibnamefont {Mochizuki}}, \bibinfo {author} {\bibfnamefont
  {N.}~\bibnamefont {Furukawa}}, \ and\ \bibinfo {author} {\bibfnamefont
  {Y.}~\bibnamefont {Tokura}},\ }\href@noop {} {\bibfield  {journal} {\bibinfo
  {journal} {Phys. Rev. Lett.},\ }\textbf {\bibinfo {volume} {101}},\ \bibinfo
  {pages} {187201} (\bibinfo {year} {2008})}\BibitemShut {NoStop}%
\bibitem [{\citenamefont {Wilkins}\ \emph {et~al.}(2009)\citenamefont
  {Wilkins}, \citenamefont {Forrest}, \citenamefont {Beale}, \citenamefont
  {Bland}, \citenamefont {Walker}, \citenamefont {Mannix}, \citenamefont
  {Yakhou}, \citenamefont {Prabhakaran}, \citenamefont {Boothroyd},
  \citenamefont {Hill}, \citenamefont {Hatton},\ and\ \citenamefont
  {McMorrow}}]{Wilk09}%
  \BibitemOpen
  \bibfield  {author} {\bibinfo {author} {\bibfnamefont {S.~B.}\ \bibnamefont
  {Wilkins}}, \bibinfo {author} {\bibfnamefont {T.~R.}\ \bibnamefont
  {Forrest}}, \bibinfo {author} {\bibfnamefont {T.~A.~W.}\ \bibnamefont
  {Beale}}, \bibinfo {author} {\bibfnamefont {S.~R.}\ \bibnamefont {Bland}},
  \bibinfo {author} {\bibfnamefont {H.~C.}\ \bibnamefont {Walker}}, \bibinfo
  {author} {\bibfnamefont {D.}~\bibnamefont {Mannix}}, \bibinfo {author}
  {\bibfnamefont {F.}~\bibnamefont {Yakhou}}, \bibinfo {author} {\bibfnamefont
  {D.}~\bibnamefont {Prabhakaran}}, \bibinfo {author} {\bibfnamefont {A.~T.}\
  \bibnamefont {Boothroyd}}, \bibinfo {author} {\bibfnamefont {J.~P.}\
  \bibnamefont {Hill}}, \bibinfo {author} {\bibfnamefont {P.~D.}\ \bibnamefont
  {Hatton}}, \ and\ \bibinfo {author} {\bibfnamefont {D.~F.}\ \bibnamefont
  {McMorrow}},\ }\href@noop {} {\bibfield  {journal} {\bibinfo  {journal}
  {Phys. Rev. Lett.},\ }\textbf {\bibinfo {volume} {103}},\ \bibinfo {pages}
  {207602} (\bibinfo {year} {2009})}\BibitemShut {NoStop}%
\bibitem [{\citenamefont {Aliouane}\ \emph {et~al.}(2009)\citenamefont
  {Aliouane}, \citenamefont {Schmalzl}, \citenamefont {Senff}, \citenamefont
  {Maljuk}, \citenamefont {Prokes}, \citenamefont {Braden},\ and\ \citenamefont
  {Argyriou}}]{Ali09}%
  \BibitemOpen
  \bibfield  {author} {\bibinfo {author} {\bibfnamefont {N.}~\bibnamefont
  {Aliouane}}, \bibinfo {author} {\bibfnamefont {K.}~\bibnamefont {Schmalzl}},
  \bibinfo {author} {\bibfnamefont {D.}~\bibnamefont {Senff}}, \bibinfo
  {author} {\bibfnamefont {A.}~\bibnamefont {Maljuk}}, \bibinfo {author}
  {\bibfnamefont {K.}~\bibnamefont {Prokes}}, \bibinfo {author} {\bibfnamefont
  {M.}~\bibnamefont {Braden}}, \ and\ \bibinfo {author} {\bibfnamefont {D.~N.}\
  \bibnamefont {Argyriou}},\ }\href@noop {} {\bibfield  {journal} {\bibinfo
  {journal} {Phys. Rev. Lett.},\ }\textbf {\bibinfo {volume} {102}},\ \bibinfo
  {pages} {207205} (\bibinfo {year} {2009})}\BibitemShut {NoStop}%
\bibitem [{\citenamefont {Kajimoto}\ \emph {et~al.}(2009)\citenamefont
  {Kajimoto}, \citenamefont {Sagayama}, \citenamefont {Sasai}, \citenamefont
  {Fukuda}, \citenamefont {Tsutsui}, \citenamefont {Arima}, \citenamefont
  {Hirota}, \citenamefont {Mitsui}, \citenamefont {Yoshizawa}, \citenamefont
  {Baron}, \citenamefont {Yamasaki},\ and\ \citenamefont {Tokura}}]{Kaji09}%
  \BibitemOpen
  \bibfield  {author} {\bibinfo {author} {\bibfnamefont {R.}~\bibnamefont
  {Kajimoto}}, \bibinfo {author} {\bibfnamefont {H.}~\bibnamefont {Sagayama}},
  \bibinfo {author} {\bibfnamefont {K.}~\bibnamefont {Sasai}}, \bibinfo
  {author} {\bibfnamefont {T.}~\bibnamefont {Fukuda}}, \bibinfo {author}
  {\bibfnamefont {S.}~\bibnamefont {Tsutsui}}, \bibinfo {author} {\bibfnamefont
  {T.}~\bibnamefont {Arima}}, \bibinfo {author} {\bibfnamefont
  {K.}~\bibnamefont {Hirota}}, \bibinfo {author} {\bibfnamefont
  {Y.}~\bibnamefont {Mitsui}}, \bibinfo {author} {\bibfnamefont
  {H.}~\bibnamefont {Yoshizawa}}, \bibinfo {author} {\bibfnamefont {A.~Q.~R.}\
  \bibnamefont {Baron}}, \bibinfo {author} {\bibfnamefont {Y.}~\bibnamefont
  {Yamasaki}}, \ and\ \bibinfo {author} {\bibfnamefont {Y.}~\bibnamefont
  {Tokura}},\ }\href@noop {} {\bibfield  {journal} {\bibinfo  {journal} {Phys.
  Rev. Lett.},\ }\textbf {\bibinfo {volume} {102}},\ \bibinfo {pages} {247602}
  (\bibinfo {year} {2009})}\BibitemShut {NoStop}%
\bibitem [{\citenamefont {Fabrizi}\ \emph {et~al.}(2009)\citenamefont
  {Fabrizi}, \citenamefont {Walker}, \citenamefont {Paolasini}, \citenamefont
  {de~Bergevin}, \citenamefont {Boothroyd}, \citenamefont {Prabhakaran},\ and\
  \citenamefont {McMorrow}}]{Fab09}%
  \BibitemOpen
  \bibfield  {author} {\bibinfo {author} {\bibfnamefont {F.}~\bibnamefont
  {Fabrizi}}, \bibinfo {author} {\bibfnamefont {H.~C.}\ \bibnamefont {Walker}},
  \bibinfo {author} {\bibfnamefont {L.}~\bibnamefont {Paolasini}}, \bibinfo
  {author} {\bibfnamefont {F.}~\bibnamefont {de~Bergevin}}, \bibinfo {author}
  {\bibfnamefont {A.~T.}\ \bibnamefont {Boothroyd}}, \bibinfo {author}
  {\bibfnamefont {D.}~\bibnamefont {Prabhakaran}}, \ and\ \bibinfo {author}
  {\bibfnamefont {D.~F.}\ \bibnamefont {McMorrow}},\ }\href@noop {} {\bibfield
  {journal} {\bibinfo  {journal} {Phys. Rev. Lett.},\ }\textbf {\bibinfo
  {volume} {102}},\ \bibinfo {pages} {237205} (\bibinfo {year}
  {2009})}\BibitemShut {NoStop}%
\bibitem [{\citenamefont {Shuvaev}\ \emph {et~al.}(2010)\citenamefont
  {Shuvaev}, \citenamefont {Travkin}, \citenamefont {Ivanov}, \citenamefont
  {Mukhin},\ and\ \citenamefont {Pimenov}}]{Shu10}%
  \BibitemOpen
  \bibfield  {author} {\bibinfo {author} {\bibfnamefont {A.~M.}\ \bibnamefont
  {Shuvaev}}, \bibinfo {author} {\bibfnamefont {V.~D.}\ \bibnamefont
  {Travkin}}, \bibinfo {author} {\bibfnamefont {V.~Y.}\ \bibnamefont {Ivanov}},
  \bibinfo {author} {\bibfnamefont {A.~A.}\ \bibnamefont {Mukhin}}, \ and\
  \bibinfo {author} {\bibfnamefont {A.}~\bibnamefont {Pimenov}},\ }\href@noop
  {} {\bibfield  {journal} {\bibinfo  {journal} {Phys. Rev. Lett.},\ }\textbf
  {\bibinfo {volume} {104}},\ \bibinfo {pages} {097202} (\bibinfo {year}
  {2010})}\BibitemShut {NoStop}%
\bibitem [{\citenamefont {Lautenschl{\"a}ger}\ \emph
  {et~al.}(1993)\citenamefont {Lautenschl{\"a}ger}, \citenamefont {Weitzel},
  \citenamefont {Vogt}, \citenamefont {Hock}, \citenamefont {B{\"o}hm},
  \citenamefont {Bonnet},\ and\ \citenamefont {Fuess}}]{Lau93}%
  \BibitemOpen
  \bibfield  {author} {\bibinfo {author} {\bibfnamefont {G.}~\bibnamefont
  {Lautenschl{\"a}ger}}, \bibinfo {author} {\bibfnamefont {H.}~\bibnamefont
  {Weitzel}}, \bibinfo {author} {\bibfnamefont {T.}~\bibnamefont {Vogt}},
  \bibinfo {author} {\bibfnamefont {R.}~\bibnamefont {Hock}}, \bibinfo {author}
  {\bibfnamefont {A.}~\bibnamefont {B{\"o}hm}}, \bibinfo {author}
  {\bibfnamefont {M.}~\bibnamefont {Bonnet}}, \ and\ \bibinfo {author}
  {\bibfnamefont {H.}~\bibnamefont {Fuess}},\ }\href@noop {} {\bibfield
  {journal} {\bibinfo  {journal} {Phys. Rev. B},\ }\textbf {\bibinfo {volume}
  {48}},\ \bibinfo {pages} {6087} (\bibinfo {year} {1993})}\BibitemShut
  {NoStop}%
\bibitem [{\citenamefont {Taniguchi}\ \emph {et~al.}(2008)\citenamefont
  {Taniguchi}, \citenamefont {Abe}, \citenamefont {Umetsu}, \citenamefont
  {Katori},\ and\ \citenamefont {Arima}}]{Tan08a}%
  \BibitemOpen
  \bibfield  {author} {\bibinfo {author} {\bibfnamefont {K.}~\bibnamefont
  {Taniguchi}}, \bibinfo {author} {\bibfnamefont {N.}~\bibnamefont {Abe}},
  \bibinfo {author} {\bibfnamefont {H.}~\bibnamefont {Umetsu}}, \bibinfo
  {author} {\bibfnamefont {H.~A.}\ \bibnamefont {Katori}}, \ and\ \bibinfo
  {author} {\bibfnamefont {T.}~\bibnamefont {Arima}},\ }\href@noop {}
  {\bibfield  {journal} {\bibinfo  {journal} {Phys. Rev. Lett.},\ }\textbf
  {\bibinfo {volume} {101}},\ \bibinfo {pages} {207205} (\bibinfo {year}
  {2008})}\BibitemShut {NoStop}%
\bibitem [{\citenamefont {Taniguchi}\ \emph {et~al.}(2009)\citenamefont
  {Taniguchi}, \citenamefont {Abe}, \citenamefont {Ohtani},\ and\ \citenamefont
  {Arima}}]{Tan09}%
  \BibitemOpen
  \bibfield  {author} {\bibinfo {author} {\bibfnamefont {K.}~\bibnamefont
  {Taniguchi}}, \bibinfo {author} {\bibfnamefont {N.}~\bibnamefont {Abe}},
  \bibinfo {author} {\bibfnamefont {S.}~\bibnamefont {Ohtani}}, \ and\ \bibinfo
  {author} {\bibfnamefont {T.}~\bibnamefont {Arima}},\ }\href@noop {}
  {\bibfield  {journal} {\bibinfo  {journal} {Phys. Rev. Lett.},\ }\textbf
  {\bibinfo {volume} {102}},\ \bibinfo {pages} {147201} (\bibinfo {year}
  {2009})}\BibitemShut {NoStop}%
\bibitem [{\citenamefont {Shanavas}\ \emph {et~al.}(2010)\citenamefont
  {Shanavas}, \citenamefont {Choudhury}, \citenamefont {Dasgupta},
  \citenamefont {Sharma},\ and\ \citenamefont {Sarma}}]{Sha10}%
  \BibitemOpen
  \bibfield  {author} {\bibinfo {author} {\bibfnamefont {K.~V.}\ \bibnamefont
  {Shanavas}}, \bibinfo {author} {\bibfnamefont {D.}~\bibnamefont {Choudhury}},
  \bibinfo {author} {\bibfnamefont {I.}~\bibnamefont {Dasgupta}}, \bibinfo
  {author} {\bibfnamefont {S.~M.}\ \bibnamefont {Sharma}}, \ and\ \bibinfo
  {author} {\bibfnamefont {D.~D.}\ \bibnamefont {Sarma}},\ }\href@noop {}
  {\bibfield  {journal} {\bibinfo  {journal} {Phys. Rev. B},\ }\textbf
  {\bibinfo {volume} {81}},\ \bibinfo {pages} {212406} (\bibinfo {year}
  {2010})}\BibitemShut {NoStop}%
\bibitem [{\citenamefont {Klimin}\ \emph {et~al.}(2003)\citenamefont {Klimin},
  \citenamefont {Popova}, \citenamefont {Mavrin}, \citenamefont {van
  Loosdrecht}, \citenamefont {Svistov}, \citenamefont {Smirnov}, \citenamefont
  {Prozorova}, \citenamefont {von Nidda}, \citenamefont {Seidov}, \citenamefont
  {Loidl}, \citenamefont {Shapiro},\ and\ \citenamefont
  {Demianets}}]{Klimin03}%
  \BibitemOpen
  \bibfield  {author} {\bibinfo {author} {\bibfnamefont {S.~A.}\ \bibnamefont
  {Klimin}}, \bibinfo {author} {\bibfnamefont {M.~N.}\ \bibnamefont {Popova}},
  \bibinfo {author} {\bibfnamefont {B.~N.}\ \bibnamefont {Mavrin}}, \bibinfo
  {author} {\bibfnamefont {P.~H.~M.}\ \bibnamefont {van Loosdrecht}}, \bibinfo
  {author} {\bibfnamefont {L.~E.}\ \bibnamefont {Svistov}}, \bibinfo {author}
  {\bibfnamefont {A.~I.}\ \bibnamefont {Smirnov}}, \bibinfo {author}
  {\bibfnamefont {L.~A.}\ \bibnamefont {Prozorova}}, \bibinfo {author}
  {\bibfnamefont {H.-A.~K.}\ \bibnamefont {von Nidda}}, \bibinfo {author}
  {\bibfnamefont {Z.}~\bibnamefont {Seidov}}, \bibinfo {author} {\bibfnamefont
  {A.}~\bibnamefont {Loidl}}, \bibinfo {author} {\bibfnamefont {A.~Y.}\
  \bibnamefont {Shapiro}}, \ and\ \bibinfo {author} {\bibfnamefont {L.~N.}\
  \bibnamefont {Demianets}},\ }\href@noop {} {\bibfield  {journal} {\bibinfo
  {journal} {Phys. Rev. B},\ }\textbf {\bibinfo {volume} {68}},\ \bibinfo
  {pages} {174408} (\bibinfo {year} {2003})}\BibitemShut {NoStop}%
\bibitem [{\citenamefont {Kenzelmann}\ \emph {et~al.}(2007)\citenamefont
  {Kenzelmann}, \citenamefont {Lawes}, \citenamefont {Harris}, \citenamefont
  {Gasparovic}, \citenamefont {Broholm}, \citenamefont {Ramirez}, \citenamefont
  {Jorge}, \citenamefont {Jaime}, \citenamefont {Park}, \citenamefont {Huang},
  \citenamefont {Shapiro},\ and\ \citenamefont {Demianets}}]{Ken07}%
  \BibitemOpen
  \bibfield  {author} {\bibinfo {author} {\bibfnamefont {M.}~\bibnamefont
  {Kenzelmann}}, \bibinfo {author} {\bibfnamefont {G.}~\bibnamefont {Lawes}},
  \bibinfo {author} {\bibfnamefont {A.~B.}\ \bibnamefont {Harris}}, \bibinfo
  {author} {\bibfnamefont {G.}~\bibnamefont {Gasparovic}}, \bibinfo {author}
  {\bibfnamefont {C.}~\bibnamefont {Broholm}}, \bibinfo {author} {\bibfnamefont
  {A.~P.}\ \bibnamefont {Ramirez}}, \bibinfo {author} {\bibfnamefont {G.~A.}\
  \bibnamefont {Jorge}}, \bibinfo {author} {\bibfnamefont {M.}~\bibnamefont
  {Jaime}}, \bibinfo {author} {\bibfnamefont {S.}~\bibnamefont {Park}},
  \bibinfo {author} {\bibfnamefont {Q.}~\bibnamefont {Huang}}, \bibinfo
  {author} {\bibfnamefont {A.~Y.}\ \bibnamefont {Shapiro}}, \ and\ \bibinfo
  {author} {\bibfnamefont {L.~A.}\ \bibnamefont {Demianets}},\ }\href@noop {}
  {\bibfield  {journal} {\bibinfo  {journal} {Phys. Rev. Lett.},\ }\textbf
  {\bibinfo {volume} {98}},\ \bibinfo {pages} {267205} (\bibinfo {year}
  {2007})}\BibitemShut {NoStop}%
\bibitem [{\citenamefont {Masuda}\ \emph {et~al.}(2004)\citenamefont {Masuda},
  \citenamefont {Zheludev}, \citenamefont {Bush}, \citenamefont {Markina},\
  and\ \citenamefont {Vasiliev}}]{Ma04}%
  \BibitemOpen
  \bibfield  {author} {\bibinfo {author} {\bibfnamefont {T.}~\bibnamefont
  {Masuda}}, \bibinfo {author} {\bibfnamefont {A.}~\bibnamefont {Zheludev}},
  \bibinfo {author} {\bibfnamefont {A.}~\bibnamefont {Bush}}, \bibinfo {author}
  {\bibfnamefont {M.}~\bibnamefont {Markina}}, \ and\ \bibinfo {author}
  {\bibfnamefont {A.}~\bibnamefont {Vasiliev}},\ }\href@noop {} {\bibfield
  {journal} {\bibinfo  {journal} {Phys. Rev. Lett.},\ }\textbf {\bibinfo
  {volume} {92}},\ \bibinfo {pages} {177201} (\bibinfo {year}
  {2004})}\BibitemShut {NoStop}%
\bibitem [{\citenamefont {Masuda}\ \emph {et~al.}(2005)\citenamefont {Masuda},
  \citenamefont {Zheludev}, \citenamefont {Roessli}, \citenamefont {Bush},
  \citenamefont {Markina},\ and\ \citenamefont {Vasiliev}}]{Ma05}%
  \BibitemOpen
  \bibfield  {author} {\bibinfo {author} {\bibfnamefont {T.}~\bibnamefont
  {Masuda}}, \bibinfo {author} {\bibfnamefont {A.}~\bibnamefont {Zheludev}},
  \bibinfo {author} {\bibfnamefont {B.}~\bibnamefont {Roessli}}, \bibinfo
  {author} {\bibfnamefont {A.}~\bibnamefont {Bush}}, \bibinfo {author}
  {\bibfnamefont {M.}~\bibnamefont {Markina}}, \ and\ \bibinfo {author}
  {\bibfnamefont {A.}~\bibnamefont {Vasiliev}},\ }\href@noop {} {\bibfield
  {journal} {\bibinfo  {journal} {Phys. Rev. B},\ }\textbf {\bibinfo {volume}
  {72}},\ \bibinfo {pages} {014405} (\bibinfo {year} {2005})}\BibitemShut
  {NoStop}%
\bibitem [{\citenamefont {Xiang}\ and\ \citenamefont
  {Whangbo}(2007)}]{Xiang07}%
  \BibitemOpen
  \bibfield  {author} {\bibinfo {author} {\bibfnamefont {H.~J.}\ \bibnamefont
  {Xiang}}\ and\ \bibinfo {author} {\bibfnamefont {M.-H.}\ \bibnamefont
  {Whangbo}},\ }\href@noop {} {\bibfield  {journal} {\bibinfo  {journal} {Phys.
  Rev. Lett.},\ }\textbf {\bibinfo {volume} {99}},\ \bibinfo {pages} {257203}
  (\bibinfo {year} {2007})}\BibitemShut {NoStop}%
\bibitem [{\citenamefont {Huang}\ \emph {et~al.}(2008)\citenamefont {Huang},
  \citenamefont {Huang}, \citenamefont {Okamoto}, \citenamefont {Mou},
  \citenamefont {Wu}, \citenamefont {Yeh}, \citenamefont {Chen}, \citenamefont
  {Wu}, \citenamefont {Hsu}, \citenamefont {Chou},\ and\ \citenamefont
  {Chen}}]{Hu08}%
  \BibitemOpen
  \bibfield  {author} {\bibinfo {author} {\bibfnamefont {S.~W.}\ \bibnamefont
  {Huang}}, \bibinfo {author} {\bibfnamefont {D.~J.}\ \bibnamefont {Huang}},
  \bibinfo {author} {\bibfnamefont {J.}~\bibnamefont {Okamoto}}, \bibinfo
  {author} {\bibfnamefont {C.~Y.}\ \bibnamefont {Mou}}, \bibinfo {author}
  {\bibfnamefont {W.~B.}\ \bibnamefont {Wu}}, \bibinfo {author} {\bibfnamefont
  {K.~W.}\ \bibnamefont {Yeh}}, \bibinfo {author} {\bibfnamefont {C.~L.}\
  \bibnamefont {Chen}}, \bibinfo {author} {\bibfnamefont {M.~K.}\ \bibnamefont
  {Wu}}, \bibinfo {author} {\bibfnamefont {H.~C.}\ \bibnamefont {Hsu}},
  \bibinfo {author} {\bibfnamefont {F.~C.}\ \bibnamefont {Chou}}, \ and\
  \bibinfo {author} {\bibfnamefont {C.~T.}\ \bibnamefont {Chen}},\ }\href@noop
  {} {\bibfield  {journal} {\bibinfo  {journal} {Phys. Rev. Lett.},\ }\textbf
  {\bibinfo {volume} {101}},\ \bibinfo {pages} {077205} (\bibinfo {year}
  {2008})}\BibitemShut {NoStop}%
\bibitem [{\citenamefont {H{\"u}vonen}\ \emph {et~al.}(2009)\citenamefont
  {H{\"u}vonen}, \citenamefont {Nagel}, \citenamefont {R{\~o}{\~o}m},
  \citenamefont {Choi}, \citenamefont {Zhang}, \citenamefont {Park},\ and\
  \citenamefont {Cheong}}]{Hu09}%
  \BibitemOpen
  \bibfield  {author} {\bibinfo {author} {\bibfnamefont {D.}~\bibnamefont
  {H{\"u}vonen}}, \bibinfo {author} {\bibfnamefont {U.}~\bibnamefont {Nagel}},
  \bibinfo {author} {\bibfnamefont {T.}~\bibnamefont {R{\~o}{\~o}m}}, \bibinfo
  {author} {\bibfnamefont {Y.~J.}\ \bibnamefont {Choi}}, \bibinfo {author}
  {\bibfnamefont {C.~L.}\ \bibnamefont {Zhang}}, \bibinfo {author}
  {\bibfnamefont {S.}~\bibnamefont {Park}}, \ and\ \bibinfo {author}
  {\bibfnamefont {S.-W.}\ \bibnamefont {Cheong}},\ }\href@noop {} {\bibfield
  {journal} {\bibinfo  {journal} {Phys. Rev. B},\ }\textbf {\bibinfo {volume}
  {80}},\ \bibinfo {pages} {100402(R)} (\bibinfo {year} {2009})}\BibitemShut
  {NoStop}%
\bibitem [{\citenamefont {Seki}\ \emph
  {et~al.}(2008){\natexlab{a}}\citenamefont {Seki}, \citenamefont {Yamasaki},
  \citenamefont {Soda}, \citenamefont {Matsuura}, \citenamefont {Hirota},\ and\
  \citenamefont {Tokura}}]{Seki08b}%
  \BibitemOpen
  \bibfield  {author} {\bibinfo {author} {\bibfnamefont {S.}~\bibnamefont
  {Seki}}, \bibinfo {author} {\bibfnamefont {Y.}~\bibnamefont {Yamasaki}},
  \bibinfo {author} {\bibfnamefont {M.}~\bibnamefont {Soda}}, \bibinfo {author}
  {\bibfnamefont {M.}~\bibnamefont {Matsuura}}, \bibinfo {author}
  {\bibfnamefont {K.}~\bibnamefont {Hirota}}, \ and\ \bibinfo {author}
  {\bibfnamefont {Y.}~\bibnamefont {Tokura}},\ }\href@noop {} {\bibfield
  {journal} {\bibinfo  {journal} {Phys. Rev. Lett.},\ }\textbf {\bibinfo
  {volume} {100}},\ \bibinfo {pages} {127201} (\bibinfo {year}
  {2008}{\natexlab{a}})}\BibitemShut {NoStop}%
\bibitem [{\citenamefont {Lawes}\ \emph {et~al.}(2004)\citenamefont {Lawes},
  \citenamefont {Kenzelmann}, \citenamefont {Rogado}, \citenamefont {Kim},
  \citenamefont {Jorge}, \citenamefont {Cava}, \citenamefont {Aharony},
  \citenamefont {Entin-Wohlman}, \citenamefont {Harris}, \citenamefont
  {Yildirim}, \citenamefont {Huang}, \citenamefont {Park}, \citenamefont
  {Broholm},\ and\ \citenamefont {Ramirez}}]{Lawes04}%
  \BibitemOpen
  \bibfield  {author} {\bibinfo {author} {\bibfnamefont {G.}~\bibnamefont
  {Lawes}}, \bibinfo {author} {\bibfnamefont {M.}~\bibnamefont {Kenzelmann}},
  \bibinfo {author} {\bibfnamefont {N.}~\bibnamefont {Rogado}}, \bibinfo
  {author} {\bibfnamefont {K.~H.}\ \bibnamefont {Kim}}, \bibinfo {author}
  {\bibfnamefont {G.~A.}\ \bibnamefont {Jorge}}, \bibinfo {author}
  {\bibfnamefont {R.~J.}\ \bibnamefont {Cava}}, \bibinfo {author}
  {\bibfnamefont {A.}~\bibnamefont {Aharony}}, \bibinfo {author} {\bibfnamefont
  {O.}~\bibnamefont {Entin-Wohlman}}, \bibinfo {author} {\bibfnamefont {A.~B.}\
  \bibnamefont {Harris}}, \bibinfo {author} {\bibfnamefont {T.}~\bibnamefont
  {Yildirim}}, \bibinfo {author} {\bibfnamefont {Q.~Z.}\ \bibnamefont {Huang}},
  \bibinfo {author} {\bibfnamefont {S.}~\bibnamefont {Park}}, \bibinfo {author}
  {\bibfnamefont {C.}~\bibnamefont {Broholm}}, \ and\ \bibinfo {author}
  {\bibfnamefont {A.~P.}\ \bibnamefont {Ramirez}},\ }\href@noop {} {\bibfield
  {journal} {\bibinfo  {journal} {Phys. Rev. Lett.},\ }\textbf {\bibinfo
  {volume} {93}},\ \bibinfo {pages} {247201} (\bibinfo {year}
  {2004})}\BibitemShut {NoStop}%
\bibitem [{\citenamefont {Lawes}\ \emph {et~al.}(2005)\citenamefont {Lawes},
  \citenamefont {Harris}, \citenamefont {Kimura}, \citenamefont {Rogado},
  \citenamefont {Cava}, \citenamefont {Aharony}, \citenamefont {Entin-Wohlman},
  \citenamefont {Yildrim}, \citenamefont {Kenzelmann}, \citenamefont
  {Broholm},\ and\ \citenamefont {Ramirez}}]{Lawes05}%
  \BibitemOpen
  \bibfield  {author} {\bibinfo {author} {\bibfnamefont {G.}~\bibnamefont
  {Lawes}}, \bibinfo {author} {\bibfnamefont {A.~B.}\ \bibnamefont {Harris}},
  \bibinfo {author} {\bibfnamefont {T.}~\bibnamefont {Kimura}}, \bibinfo
  {author} {\bibfnamefont {N.}~\bibnamefont {Rogado}}, \bibinfo {author}
  {\bibfnamefont {R.~J.}\ \bibnamefont {Cava}}, \bibinfo {author}
  {\bibfnamefont {A.}~\bibnamefont {Aharony}}, \bibinfo {author} {\bibfnamefont
  {O.}~\bibnamefont {Entin-Wohlman}}, \bibinfo {author} {\bibfnamefont
  {T.}~\bibnamefont {Yildrim}}, \bibinfo {author} {\bibfnamefont
  {M.}~\bibnamefont {Kenzelmann}}, \bibinfo {author} {\bibfnamefont
  {C.}~\bibnamefont {Broholm}}, \ and\ \bibinfo {author} {\bibfnamefont
  {A.~P.}\ \bibnamefont {Ramirez}},\ }\href@noop {} {\bibfield  {journal}
  {\bibinfo  {journal} {Phys. Rev. Lett.},\ }\textbf {\bibinfo {volume} {95}},\
  \bibinfo {pages} {087205} (\bibinfo {year} {2005})}\BibitemShut {NoStop}%
\bibitem [{\citenamefont {Harris}\ \emph {et~al.}(2006)\citenamefont {Harris},
  \citenamefont {Yildirim}, \citenamefont {Aharony},\ and\ \citenamefont
  {Entin-Wohlman}}]{Har06}%
  \BibitemOpen
  \bibfield  {author} {\bibinfo {author} {\bibfnamefont {A.~B.}\ \bibnamefont
  {Harris}}, \bibinfo {author} {\bibfnamefont {T.}~\bibnamefont {Yildirim}},
  \bibinfo {author} {\bibfnamefont {A.}~\bibnamefont {Aharony}}, \ and\
  \bibinfo {author} {\bibfnamefont {O.}~\bibnamefont {Entin-Wohlman}},\
  }\href@noop {} {\bibfield  {journal} {\bibinfo  {journal} {Phys. Rev. B},\
  }\textbf {\bibinfo {volume} {73}},\ \bibinfo {pages} {184433} (\bibinfo
  {year} {2006})}\BibitemShut {NoStop}%
\bibitem [{\citenamefont {Uhrmacher}\ \emph {et~al.}(1996)\citenamefont
  {Uhrmacher}, \citenamefont {Attili}, \citenamefont {Lieb}, \citenamefont
  {Winzer},\ and\ \citenamefont {Mekata}}]{Uhr96}%
  \BibitemOpen
  \bibfield  {author} {\bibinfo {author} {\bibfnamefont {M.}~\bibnamefont
  {Uhrmacher}}, \bibinfo {author} {\bibfnamefont {R.~N.}\ \bibnamefont
  {Attili}}, \bibinfo {author} {\bibfnamefont {K.~P.}\ \bibnamefont {Lieb}},
  \bibinfo {author} {\bibfnamefont {K.}~\bibnamefont {Winzer}}, \ and\ \bibinfo
  {author} {\bibfnamefont {M.}~\bibnamefont {Mekata}},\ }\href@noop {}
  {\bibfield  {journal} {\bibinfo  {journal} {Phys. Rev. Lett.},\ }\textbf
  {\bibinfo {volume} {76}},\ \bibinfo {pages} {4829} (\bibinfo {year}
  {1996})}\BibitemShut {NoStop}%
\bibitem [{\citenamefont {Ye}\ \emph {et~al.}(2007)\citenamefont {Ye},
  \citenamefont {Fernandez-Baca}, \citenamefont {Fishman}, \citenamefont {Ren},
  \citenamefont {Kang}, \citenamefont {Qiu},\ and\ \citenamefont
  {Kimura}}]{Ye07}%
  \BibitemOpen
  \bibfield  {author} {\bibinfo {author} {\bibfnamefont {F.}~\bibnamefont
  {Ye}}, \bibinfo {author} {\bibfnamefont {J.~A.}\ \bibnamefont
  {Fernandez-Baca}}, \bibinfo {author} {\bibfnamefont {R.~S.}\ \bibnamefont
  {Fishman}}, \bibinfo {author} {\bibfnamefont {Y.}~\bibnamefont {Ren}},
  \bibinfo {author} {\bibfnamefont {H.~J.}\ \bibnamefont {Kang}}, \bibinfo
  {author} {\bibfnamefont {Y.}~\bibnamefont {Qiu}}, \ and\ \bibinfo {author}
  {\bibfnamefont {T.}~\bibnamefont {Kimura}},\ }\href@noop {} {\bibfield
  {journal} {\bibinfo  {journal} {Phys. Rev. Lett.},\ }\textbf {\bibinfo
  {volume} {99}},\ \bibinfo {pages} {157201} (\bibinfo {year}
  {2007})}\BibitemShut {NoStop}%
\bibitem [{\citenamefont {Wang}\ and\ \citenamefont
  {Vishwanath}(2008)}]{Wang08}%
  \BibitemOpen
  \bibfield  {author} {\bibinfo {author} {\bibfnamefont {F.}~\bibnamefont
  {Wang}}\ and\ \bibinfo {author} {\bibfnamefont {A.}~\bibnamefont
  {Vishwanath}},\ }\href@noop {} {\bibfield  {journal} {\bibinfo  {journal}
  {Phys. Rev. Lett.},\ }\textbf {\bibinfo {volume} {100}},\ \bibinfo {pages}
  {077201} (\bibinfo {year} {2008})}\BibitemShut {NoStop}%
\bibitem [{\citenamefont {Haraldsen}\ \emph {et~al.}(2009)\citenamefont
  {Haraldsen}, \citenamefont {Swanson}, \citenamefont {Alvarez},\ and\
  \citenamefont {Fishman}}]{Har09a}%
  \BibitemOpen
  \bibfield  {author} {\bibinfo {author} {\bibfnamefont {J.~T.}\ \bibnamefont
  {Haraldsen}}, \bibinfo {author} {\bibfnamefont {M.}~\bibnamefont {Swanson}},
  \bibinfo {author} {\bibfnamefont {G.}~\bibnamefont {Alvarez}}, \ and\
  \bibinfo {author} {\bibfnamefont {R.~S.}\ \bibnamefont {Fishman}},\
  }\href@noop {} {\bibfield  {journal} {\bibinfo  {journal} {Phys. Rev.
  Lett.},\ }\textbf {\bibinfo {volume} {102}},\ \bibinfo {pages} {237204}
  (\bibinfo {year} {2009})}\BibitemShut {NoStop}%
\bibitem [{\citenamefont {Nakajima}\ \emph {et~al.}(2009)\citenamefont
  {Nakajima}, \citenamefont {Mitsuda}, \citenamefont {Takahashi}, \citenamefont
  {Yamano}, \citenamefont {Masuda}, \citenamefont {Yamazaki}, \citenamefont
  {Prokes}, \citenamefont {Kiefer}, \citenamefont {Gerischer}, \citenamefont
  {Terada}, \citenamefont {Kitazawa}, \citenamefont {Matsuda}, \citenamefont
  {Kakurai}, \citenamefont {Kimura}, \citenamefont {Noda}, \citenamefont
  {Soda}, \citenamefont {Matsuura},\ and\ \citenamefont {Hirota}}]{Nak09}%
  \BibitemOpen
  \bibfield  {author} {\bibinfo {author} {\bibfnamefont {T.}~\bibnamefont
  {Nakajima}}, \bibinfo {author} {\bibfnamefont {S.}~\bibnamefont {Mitsuda}},
  \bibinfo {author} {\bibfnamefont {K.}~\bibnamefont {Takahashi}}, \bibinfo
  {author} {\bibfnamefont {M.}~\bibnamefont {Yamano}}, \bibinfo {author}
  {\bibfnamefont {K.}~\bibnamefont {Masuda}}, \bibinfo {author} {\bibfnamefont
  {H.}~\bibnamefont {Yamazaki}}, \bibinfo {author} {\bibfnamefont
  {K.}~\bibnamefont {Prokes}}, \bibinfo {author} {\bibfnamefont
  {K.}~\bibnamefont {Kiefer}}, \bibinfo {author} {\bibfnamefont
  {S.}~\bibnamefont {Gerischer}}, \bibinfo {author} {\bibfnamefont
  {N.}~\bibnamefont {Terada}}, \bibinfo {author} {\bibfnamefont
  {H.}~\bibnamefont {Kitazawa}}, \bibinfo {author} {\bibfnamefont
  {M.}~\bibnamefont {Matsuda}}, \bibinfo {author} {\bibfnamefont
  {K.}~\bibnamefont {Kakurai}}, \bibinfo {author} {\bibfnamefont
  {H.}~\bibnamefont {Kimura}}, \bibinfo {author} {\bibfnamefont
  {Y.}~\bibnamefont {Noda}}, \bibinfo {author} {\bibfnamefont {M.}~\bibnamefont
  {Soda}}, \bibinfo {author} {\bibfnamefont {M.}~\bibnamefont {Matsuura}}, \
  and\ \bibinfo {author} {\bibfnamefont {K.}~\bibnamefont {Hirota}},\
  }\href@noop {} {\bibfield  {journal} {\bibinfo  {journal} {Phys. Rev. B},\
  }\textbf {\bibinfo {volume} {79}},\ \bibinfo {pages} {214423} (\bibinfo
  {year} {2009})}\BibitemShut {NoStop}%
\bibitem [{\citenamefont {Quirion}\ \emph {et~al.}(2009)\citenamefont
  {Quirion}, \citenamefont {Plumer}, \citenamefont {Petrenko}, \citenamefont
  {Balakrishnan},\ and\ \citenamefont {Proust}}]{Qui09}%
  \BibitemOpen
  \bibfield  {author} {\bibinfo {author} {\bibfnamefont {G.}~\bibnamefont
  {Quirion}}, \bibinfo {author} {\bibfnamefont {M.~L.}\ \bibnamefont {Plumer}},
  \bibinfo {author} {\bibfnamefont {O.~A.}\ \bibnamefont {Petrenko}}, \bibinfo
  {author} {\bibfnamefont {G.}~\bibnamefont {Balakrishnan}}, \ and\ \bibinfo
  {author} {\bibfnamefont {C.}~\bibnamefont {Proust}},\ }\href@noop {}
  {\bibfield  {journal} {\bibinfo  {journal} {Phys. Rev. B},\ }\textbf
  {\bibinfo {volume} {80}},\ \bibinfo {pages} {064420} (\bibinfo {year}
  {2009})}\BibitemShut {NoStop}%
\bibitem [{\citenamefont {Lummen}\ \emph {et~al.}(2010)\citenamefont {Lummen},
  \citenamefont {Strohm}, \citenamefont {Rakoto},\ and\ \citenamefont {van
  Loosdrecht}}]{Lum10}%
  \BibitemOpen
  \bibfield  {author} {\bibinfo {author} {\bibfnamefont {T.~T.~A.}\
  \bibnamefont {Lummen}}, \bibinfo {author} {\bibfnamefont {C.}~\bibnamefont
  {Strohm}}, \bibinfo {author} {\bibfnamefont {H.}~\bibnamefont {Rakoto}}, \
  and\ \bibinfo {author} {\bibfnamefont {P.~H.~M.}\ \bibnamefont {van
  Loosdrecht}},\ }\href@noop {} {\bibfield  {journal} {\bibinfo  {journal}
  {Phys. Rev. B},\ }\textbf {\bibinfo {volume} {81}},\ \bibinfo {pages}
  {224420} (\bibinfo {year} {2010})}\BibitemShut {NoStop}%
\bibitem [{\citenamefont {Seki}\ \emph
  {et~al.}(2008){\natexlab{b}}\citenamefont {Seki}, \citenamefont {Onose},\
  and\ \citenamefont {Tokura}}]{Seki08}%
  \BibitemOpen
  \bibfield  {author} {\bibinfo {author} {\bibfnamefont {S.}~\bibnamefont
  {Seki}}, \bibinfo {author} {\bibfnamefont {Y.}~\bibnamefont {Onose}}, \ and\
  \bibinfo {author} {\bibfnamefont {Y.}~\bibnamefont {Tokura}},\ }\Doi
  {10.1103/PhysRevLett.101.067204} {\bibfield  {journal} {\bibinfo  {journal}
  {Phys. Rev. Lett.},\ }\textbf {\bibinfo {volume} {101}},\ \bibinfo {pages}
  {067204} (\bibinfo {year} {2008}{\natexlab{b}})}\BibitemShut {NoStop}%
\bibitem [{\citenamefont {Kimura}\ \emph
  {et~al.}(2009){\natexlab{a}}\citenamefont {Kimura}, \citenamefont {Nakamura},
  \citenamefont {Kimura}, \citenamefont {Hagiwara},\ and\ \citenamefont
  {Kimura}}]{Kim09}%
  \BibitemOpen
  \bibfield  {author} {\bibinfo {author} {\bibfnamefont {K.}~\bibnamefont
  {Kimura}}, \bibinfo {author} {\bibfnamefont {H.}~\bibnamefont {Nakamura}},
  \bibinfo {author} {\bibfnamefont {S.}~\bibnamefont {Kimura}}, \bibinfo
  {author} {\bibfnamefont {M.}~\bibnamefont {Hagiwara}}, \ and\ \bibinfo
  {author} {\bibfnamefont {T.}~\bibnamefont {Kimura}},\ }\href@noop {}
  {\bibfield  {journal} {\bibinfo  {journal} {Phys. Rev. Lett.},\ }\textbf
  {\bibinfo {volume} {103}},\ \bibinfo {pages} {107201} (\bibinfo {year}
  {2009}{\natexlab{a}})}\BibitemShut {NoStop}%
\bibitem [{\citenamefont {Yamaguchi}\ \emph {et~al.}(2010)\citenamefont
  {Yamaguchi}, \citenamefont {Ohtomo}, \citenamefont {Kimura}, \citenamefont
  {Hagiwara}, \citenamefont {Kimura}, \citenamefont {Kimura}, \citenamefont
  {Okuda},\ and\ \citenamefont {Kindo}}]{Yama10}%
  \BibitemOpen
  \bibfield  {author} {\bibinfo {author} {\bibfnamefont {H.}~\bibnamefont
  {Yamaguchi}}, \bibinfo {author} {\bibfnamefont {S.}~\bibnamefont {Ohtomo}},
  \bibinfo {author} {\bibfnamefont {S.}~\bibnamefont {Kimura}}, \bibinfo
  {author} {\bibfnamefont {M.}~\bibnamefont {Hagiwara}}, \bibinfo {author}
  {\bibfnamefont {K.}~\bibnamefont {Kimura}}, \bibinfo {author} {\bibfnamefont
  {T.}~\bibnamefont {Kimura}}, \bibinfo {author} {\bibfnamefont
  {T.}~\bibnamefont {Okuda}}, \ and\ \bibinfo {author} {\bibfnamefont
  {K.}~\bibnamefont {Kindo}},\ }\href@noop {} {\bibfield  {journal} {\bibinfo
  {journal} {Phys. Rev. B},\ }\textbf {\bibinfo {volume} {81}},\ \bibinfo
  {pages} {033104} (\bibinfo {year} {2010})}\BibitemShut {NoStop}%
\bibitem [{\citenamefont {Arnold}\ \emph {et~al.}(2009)\citenamefont {Arnold},
  \citenamefont {Payne}, \citenamefont {Bourlange}, \citenamefont {Hu},
  \citenamefont {Egdell}, \citenamefont {Piper}, \citenamefont {Colakerol},
  \citenamefont {DeMasi}, \citenamefont {Glans}, \citenamefont {Learmonth},
  \citenamefont {Smith}, \citenamefont {Guo}, \citenamefont {Scanlon},
  \citenamefont {Walsh}, \citenamefont {Morgan},\ and\ \citenamefont
  {Watson}}]{Arn09}%
  \BibitemOpen
  \bibfield  {author} {\bibinfo {author} {\bibfnamefont {T.}~\bibnamefont
  {Arnold}}, \bibinfo {author} {\bibfnamefont {D.~J.}\ \bibnamefont {Payne}},
  \bibinfo {author} {\bibfnamefont {A.}~\bibnamefont {Bourlange}}, \bibinfo
  {author} {\bibfnamefont {J.~P.}\ \bibnamefont {Hu}}, \bibinfo {author}
  {\bibfnamefont {R.~G.}\ \bibnamefont {Egdell}}, \bibinfo {author}
  {\bibfnamefont {L.~F.~J.}\ \bibnamefont {Piper}}, \bibinfo {author}
  {\bibfnamefont {L.}~\bibnamefont {Colakerol}}, \bibinfo {author}
  {\bibfnamefont {A.}~\bibnamefont {DeMasi}}, \bibinfo {author} {\bibfnamefont
  {P.-A.}\ \bibnamefont {Glans}}, \bibinfo {author} {\bibfnamefont
  {T.}~\bibnamefont {Learmonth}}, \bibinfo {author} {\bibfnamefont {K.~E.}\
  \bibnamefont {Smith}}, \bibinfo {author} {\bibfnamefont {J.}~\bibnamefont
  {Guo}}, \bibinfo {author} {\bibfnamefont {D.~O.}\ \bibnamefont {Scanlon}},
  \bibinfo {author} {\bibfnamefont {A.}~\bibnamefont {Walsh}}, \bibinfo
  {author} {\bibfnamefont {B.~J.}\ \bibnamefont {Morgan}}, \ and\ \bibinfo
  {author} {\bibfnamefont {G.~W.}\ \bibnamefont {Watson}},\ }\href@noop {}
  {\bibfield  {journal} {\bibinfo  {journal} {Phys. Rev. B},\ }\textbf
  {\bibinfo {volume} {79}},\ \bibinfo {pages} {075102} (\bibinfo {year}
  {2009})}\BibitemShut {NoStop}%
\bibitem [{\citenamefont {Shin}\ \emph {et~al.}(2009)\citenamefont {Shin},
  \citenamefont {Foord}, \citenamefont {Payne}, \citenamefont {Arnold},
  \citenamefont {Aston}, \citenamefont {Egdell}, \citenamefont {Godinho},
  \citenamefont {Scanlon}, \citenamefont {Morgan}, \citenamefont {Watson},
  \citenamefont {Mugnier}, \citenamefont {Yaicle}, \citenamefont {Rougier},
  \citenamefont {Colakerol}, \citenamefont {Glans}, \citenamefont {Piper},\
  and\ \citenamefont {Smith}}]{Shin09}%
  \BibitemOpen
  \bibfield  {author} {\bibinfo {author} {\bibfnamefont {D.}~\bibnamefont
  {Shin}}, \bibinfo {author} {\bibfnamefont {J.~S.}\ \bibnamefont {Foord}},
  \bibinfo {author} {\bibfnamefont {D.~J.}\ \bibnamefont {Payne}}, \bibinfo
  {author} {\bibfnamefont {T.}~\bibnamefont {Arnold}}, \bibinfo {author}
  {\bibfnamefont {D.~J.}\ \bibnamefont {Aston}}, \bibinfo {author}
  {\bibfnamefont {R.~G.}\ \bibnamefont {Egdell}}, \bibinfo {author}
  {\bibfnamefont {K.~G.}\ \bibnamefont {Godinho}}, \bibinfo {author}
  {\bibfnamefont {D.~O.}\ \bibnamefont {Scanlon}}, \bibinfo {author}
  {\bibfnamefont {B.~J.}\ \bibnamefont {Morgan}}, \bibinfo {author}
  {\bibfnamefont {G.~W.}\ \bibnamefont {Watson}}, \bibinfo {author}
  {\bibfnamefont {E.}~\bibnamefont {Mugnier}}, \bibinfo {author} {\bibfnamefont
  {C.}~\bibnamefont {Yaicle}}, \bibinfo {author} {\bibfnamefont
  {A.}~\bibnamefont {Rougier}}, \bibinfo {author} {\bibfnamefont
  {L.}~\bibnamefont {Colakerol}}, \bibinfo {author} {\bibfnamefont {P.~A.}\
  \bibnamefont {Glans}}, \bibinfo {author} {\bibfnamefont {L.~F.~J.}\
  \bibnamefont {Piper}}, \ and\ \bibinfo {author} {\bibfnamefont {K.~E.}\
  \bibnamefont {Smith}},\ }\href@noop {} {\bibfield  {journal} {\bibinfo
  {journal} {Phys. Rev. B},\ }\textbf {\bibinfo {volume} {80}},\ \bibinfo
  {pages} {233105} (\bibinfo {year} {2009})}\BibitemShut {NoStop}%
\bibitem [{\citenamefont {Kimura}\ \emph
  {et~al.}(2009){\natexlab{b}}\citenamefont {Kimura}, \citenamefont {Otani},
  \citenamefont {Nakamura}, \citenamefont {Wakabayashi},\ and\ \citenamefont
  {Kimura}}]{Kim09b}%
  \BibitemOpen
  \bibfield  {author} {\bibinfo {author} {\bibfnamefont {K.}~\bibnamefont
  {Kimura}}, \bibinfo {author} {\bibfnamefont {T.}~\bibnamefont {Otani}},
  \bibinfo {author} {\bibfnamefont {H.}~\bibnamefont {Nakamura}}, \bibinfo
  {author} {\bibfnamefont {Y.}~\bibnamefont {Wakabayashi}}, \ and\ \bibinfo
  {author} {\bibfnamefont {T.}~\bibnamefont {Kimura}},\ }\Doi
  {10.1143/JPSJ.78.113710} {\bibfield  {journal} {\bibinfo  {journal} {Journal
  of the Physical Society of Japan},\ }\textbf {\bibinfo {volume} {78}},\
  \bibinfo {pages} {113710} (\bibinfo {year} {2009}{\natexlab{b}})}\BibitemShut
  {NoStop}%
\bibitem [{\citenamefont {Kadowaki}\ \emph {et~al.}(1990)\citenamefont
  {Kadowaki}, \citenamefont {Kikuchi},\ and\ \citenamefont {Ajiro}}]{Kad90}%
  \BibitemOpen
  \bibfield  {author} {\bibinfo {author} {\bibfnamefont {H.}~\bibnamefont
  {Kadowaki}}, \bibinfo {author} {\bibfnamefont {H.}~\bibnamefont {Kikuchi}}, \
  and\ \bibinfo {author} {\bibfnamefont {Y.}~\bibnamefont {Ajiro}},\ }\href
  {http://stacks.iop.org/0953-8984/2/i=19/a=014} {\bibfield  {journal}
  {\bibinfo  {journal} {Journal of Physics: Condensed Matter},\ }\textbf
  {\bibinfo {volume} {2}},\ \bibinfo {pages} {4485} (\bibinfo {year}
  {1990})}\BibitemShut {NoStop}%
\bibitem [{\citenamefont {Poienar}\ \emph {et~al.}(2009)\citenamefont
  {Poienar}, \citenamefont {Damay}, \citenamefont {Martin}, \citenamefont
  {Hardy}, \citenamefont {Maignan},\ and\ \citenamefont {Andr{\'e}}}]{Poi09}%
  \BibitemOpen
  \bibfield  {author} {\bibinfo {author} {\bibfnamefont {M.}~\bibnamefont
  {Poienar}}, \bibinfo {author} {\bibfnamefont {F.}~\bibnamefont {Damay}},
  \bibinfo {author} {\bibfnamefont {C.}~\bibnamefont {Martin}}, \bibinfo
  {author} {\bibfnamefont {V.}~\bibnamefont {Hardy}}, \bibinfo {author}
  {\bibfnamefont {A.}~\bibnamefont {Maignan}}, \ and\ \bibinfo {author}
  {\bibfnamefont {G.}~\bibnamefont {Andr{\'e}}},\ }\href@noop {} {\bibfield
  {journal} {\bibinfo  {journal} {Phys. Rev. B},\ }\textbf {\bibinfo {volume}
  {79}},\ \bibinfo {pages} {014412} (\bibinfo {year} {2009})}\BibitemShut
  {NoStop}%
\bibitem [{\citenamefont {Soda}\ \emph {et~al.}(2009)\citenamefont {Soda},
  \citenamefont {Kimura}, \citenamefont {Kimura}, \citenamefont {Matsuura},\
  and\ \citenamefont {Hirota}}]{Soda09}%
  \BibitemOpen
  \bibfield  {author} {\bibinfo {author} {\bibfnamefont {M.}~\bibnamefont
  {Soda}}, \bibinfo {author} {\bibfnamefont {K.}~\bibnamefont {Kimura}},
  \bibinfo {author} {\bibfnamefont {T.}~\bibnamefont {Kimura}}, \bibinfo
  {author} {\bibfnamefont {M.}~\bibnamefont {Matsuura}}, \ and\ \bibinfo
  {author} {\bibfnamefont {K.}~\bibnamefont {Hirota}},\ }\Doi
  {10.1143/JPSJ.78.124703} {\bibfield  {journal} {\bibinfo  {journal} {Journal
  of the Physical Society of Japan},\ }\textbf {\bibinfo {volume} {78}},\
  \bibinfo {pages} {124703} (\bibinfo {year} {2009})}\BibitemShut {NoStop}%
\bibitem [{\citenamefont {Soda}\ \emph {et~al.}(2010)\citenamefont {Soda},
  \citenamefont {Kimura}, \citenamefont {Kimura},\ and\ \citenamefont
  {Hirota}}]{Soda10}%
  \BibitemOpen
  \bibfield  {author} {\bibinfo {author} {\bibfnamefont {M.}~\bibnamefont
  {Soda}}, \bibinfo {author} {\bibfnamefont {K.}~\bibnamefont {Kimura}},
  \bibinfo {author} {\bibfnamefont {T.}~\bibnamefont {Kimura}}, \ and\ \bibinfo
  {author} {\bibfnamefont {K.}~\bibnamefont {Hirota}},\ }\href@noop {}
  {\bibfield  {journal} {\bibinfo  {journal} {Phys. Rev. B},\ }\textbf
  {\bibinfo {volume} {81}},\ \bibinfo {pages} {100406(R)} (\bibinfo {year}
  {2010})}\BibitemShut {NoStop}%
\bibitem [{\citenamefont {Frontzek}\ \emph {et~al.}(tted)\citenamefont
  {Frontzek}, \citenamefont {Ehlers}, \citenamefont {Podlesnyak}, \citenamefont
  {Cao}, \citenamefont {Matsuda}, \citenamefont {Zaharko}, \citenamefont
  {Aliouane}, \citenamefont {Barilo},\ and\ \citenamefont {Shiryaev}}]{Fron11}%
  \BibitemOpen
  \bibfield  {author} {\bibinfo {author} {\bibfnamefont {M.}~\bibnamefont
  {Frontzek}}, \bibinfo {author} {\bibfnamefont {G.}~\bibnamefont {Ehlers}},
  \bibinfo {author} {\bibfnamefont {A.}~\bibnamefont {Podlesnyak}}, \bibinfo
  {author} {\bibfnamefont {H.}~\bibnamefont {Cao}}, \bibinfo {author}
  {\bibfnamefont {M.}~\bibnamefont {Matsuda}}, \bibinfo {author} {\bibfnamefont
  {O.}~\bibnamefont {Zaharko}}, \bibinfo {author} {\bibfnamefont
  {N.}~\bibnamefont {Aliouane}}, \bibinfo {author} {\bibfnamefont
  {S.}~\bibnamefont {Barilo}}, \ and\ \bibinfo {author} {\bibfnamefont
  {S.}~\bibnamefont {Shiryaev}},\ }\href@noop {} {\bibfield  {journal}
  {\bibinfo  {journal} {Phys. Rev. B}} (\bibinfo {year} {2011,
  submitted})}\BibitemShut {NoStop}%
\bibitem [{\citenamefont {Kajimoto}\ \emph {et~al.}(2010)\citenamefont
  {Kajimoto}, \citenamefont {Nakajima}, \citenamefont {Ohira-Kawamura},
  \citenamefont {Inamura}, \citenamefont {Kakurai}, \citenamefont {Arai},
  \citenamefont {Hokazono}, \citenamefont {Oozono},\ and\ \citenamefont
  {Okuda}}]{Kaj10}%
  \BibitemOpen
  \bibfield  {author} {\bibinfo {author} {\bibfnamefont {R.}~\bibnamefont
  {Kajimoto}}, \bibinfo {author} {\bibfnamefont {K.}~\bibnamefont {Nakajima}},
  \bibinfo {author} {\bibfnamefont {S.}~\bibnamefont {Ohira-Kawamura}},
  \bibinfo {author} {\bibfnamefont {Y.}~\bibnamefont {Inamura}}, \bibinfo
  {author} {\bibfnamefont {K.}~\bibnamefont {Kakurai}}, \bibinfo {author}
  {\bibfnamefont {M.}~\bibnamefont {Arai}}, \bibinfo {author} {\bibfnamefont
  {T.}~\bibnamefont {Hokazono}}, \bibinfo {author} {\bibfnamefont
  {S.}~\bibnamefont {Oozono}}, \ and\ \bibinfo {author} {\bibfnamefont
  {T.}~\bibnamefont {Okuda}},\ }\Doi {10.1143/JPSJ.79.123705} {\bibfield
  {journal} {\bibinfo  {journal} {Journal of the Physical Society of Japan},\
  }\textbf {\bibinfo {volume} {79}},\ \bibinfo {pages} {123705} (\bibinfo
  {year} {2010})}\BibitemShut {NoStop}%
\bibitem [{\citenamefont {Poienar}\ \emph {et~al.}(2010)\citenamefont
  {Poienar}, \citenamefont {Damay}, \citenamefont {Martin}, \citenamefont
  {Robert},\ and\ \citenamefont {Petit}}]{Poi10}%
  \BibitemOpen
  \bibfield  {author} {\bibinfo {author} {\bibfnamefont {M.}~\bibnamefont
  {Poienar}}, \bibinfo {author} {\bibfnamefont {F.}~\bibnamefont {Damay}},
  \bibinfo {author} {\bibfnamefont {C.}~\bibnamefont {Martin}}, \bibinfo
  {author} {\bibfnamefont {J.}~\bibnamefont {Robert}}, \ and\ \bibinfo {author}
  {\bibfnamefont {S.}~\bibnamefont {Petit}},\ }\href@noop {} {\bibfield
  {journal} {\bibinfo  {journal} {Phys. Rev. B},\ }\textbf {\bibinfo {volume}
  {81}},\ \bibinfo {pages} {104411} (\bibinfo {year} {2010})}\BibitemShut
  {NoStop}%
\bibitem [{\citenamefont {Kimura}\ \emph {et~al.}(2006)\citenamefont {Kimura},
  \citenamefont {Lashley},\ and\ \citenamefont {Ramirez}}]{Kim06}%
  \BibitemOpen
  \bibfield  {author} {\bibinfo {author} {\bibfnamefont {T.}~\bibnamefont
  {Kimura}}, \bibinfo {author} {\bibfnamefont {J.~C.}\ \bibnamefont {Lashley}},
  \ and\ \bibinfo {author} {\bibfnamefont {A.~P.}\ \bibnamefont {Ramirez}},\
  }\href@noop {} {\bibfield  {journal} {\bibinfo  {journal} {Phys. Rev. B},\
  }\textbf {\bibinfo {volume} {73}},\ \bibinfo {pages} {220401(R)} (\bibinfo
  {year} {2006})}\BibitemShut {NoStop}%
\bibitem [{\citenamefont {Seki}\ \emph {et~al.}(2007)\citenamefont {Seki},
  \citenamefont {Yamasaki}, \citenamefont {Shiomi}, \citenamefont {Iguchi},
  \citenamefont {Onose},\ and\ \citenamefont {Tokura}}]{Seki07}%
  \BibitemOpen
  \bibfield  {author} {\bibinfo {author} {\bibfnamefont {S.}~\bibnamefont
  {Seki}}, \bibinfo {author} {\bibfnamefont {Y.}~\bibnamefont {Yamasaki}},
  \bibinfo {author} {\bibfnamefont {Y.}~\bibnamefont {Shiomi}}, \bibinfo
  {author} {\bibfnamefont {S.}~\bibnamefont {Iguchi}}, \bibinfo {author}
  {\bibfnamefont {Y.}~\bibnamefont {Onose}}, \ and\ \bibinfo {author}
  {\bibfnamefont {Y.}~\bibnamefont {Tokura}},\ }\href@noop {} {\bibfield
  {journal} {\bibinfo  {journal} {Phys. Rev. B},\ }\textbf {\bibinfo {volume}
  {75}},\ \bibinfo {pages} {100403(R)} (\bibinfo {year} {2007})}\BibitemShut
  {NoStop}%
\bibitem [{\citenamefont {Haraldsen}\ and\ \citenamefont
  {Fishman}(2010)}]{Har10}%
  \BibitemOpen
  \bibfield  {author} {\bibinfo {author} {\bibfnamefont {J.~T.}\ \bibnamefont
  {Haraldsen}}\ and\ \bibinfo {author} {\bibfnamefont {R.~S.}\ \bibnamefont
  {Fishman}},\ }\Doi {10.1103/PhysRevB.82.144441} {\bibfield  {journal}
  {\bibinfo  {journal} {Phys. Rev. B},\ }\textbf {\bibinfo {volume} {82}},\
  \bibinfo {pages} {144441} (\bibinfo {year} {2010})}\BibitemShut {NoStop}%
\bibitem [{\citenamefont {Kundys}\ \emph {et~al.}(2009)\citenamefont {Kundys},
  \citenamefont {Maignan}, \citenamefont {Pelloquin},\ and\ \citenamefont
  {Simon}}]{Kun09}%
  \BibitemOpen
  \bibfield  {author} {\bibinfo {author} {\bibfnamefont {B.}~\bibnamefont
  {Kundys}}, \bibinfo {author} {\bibfnamefont {A.}~\bibnamefont {Maignan}},
  \bibinfo {author} {\bibfnamefont {D.}~\bibnamefont {Pelloquin}}, \ and\
  \bibinfo {author} {\bibfnamefont {C.}~\bibnamefont {Simon}},\ }\Doi {DOI:
  10.1016/j.solidstatesciences.2009.02.008} {\bibfield  {journal} {\bibinfo
  {journal} {Solid State Sciences},\ }\textbf {\bibinfo {volume} {11}},\
  \bibinfo {pages} {1035 } (\bibinfo {year} {2009})},\ ISSN \bibinfo {issn}
  {1293-2558},\ \bibinfo {note} {e-MRS symposium N and R}\BibitemShut {NoStop}%
\bibitem [{\citenamefont {Nakajima}\ \emph {et~al.}(2007)\citenamefont
  {Nakajima}, \citenamefont {Mitsuda}, \citenamefont {Kanetsuki}, \citenamefont
  {Prokes}, \citenamefont {Podlesnyak}, \citenamefont {Kimura},\ and\
  \citenamefont {Noda}}]{Nak07}%
  \BibitemOpen
  \bibfield  {author} {\bibinfo {author} {\bibfnamefont {T.}~\bibnamefont
  {Nakajima}}, \bibinfo {author} {\bibfnamefont {S.}~\bibnamefont {Mitsuda}},
  \bibinfo {author} {\bibfnamefont {S.}~\bibnamefont {Kanetsuki}}, \bibinfo
  {author} {\bibfnamefont {K.}~\bibnamefont {Prokes}}, \bibinfo {author}
  {\bibfnamefont {A.}~\bibnamefont {Podlesnyak}}, \bibinfo {author}
  {\bibfnamefont {H.}~\bibnamefont {Kimura}}, \ and\ \bibinfo {author}
  {\bibfnamefont {Y.}~\bibnamefont {Noda}},\ }\Doi {10.1143/JPSJ.76.043709}
  {\bibfield  {journal} {\bibinfo  {journal} {Journal of the Physical Society
  of Japan},\ }\textbf {\bibinfo {volume} {76}},\ \bibinfo {pages} {043709}
  (\bibinfo {year} {2007})}\BibitemShut {NoStop}%
\bibitem [{\citenamefont {Kimura}\ \emph {et~al.}(2008)\citenamefont {Kimura},
  \citenamefont {Nakamura}, \citenamefont {Ohgushi},\ and\ \citenamefont
  {Kimura}}]{Kim08}%
  \BibitemOpen
  \bibfield  {author} {\bibinfo {author} {\bibfnamefont {K.}~\bibnamefont
  {Kimura}}, \bibinfo {author} {\bibfnamefont {H.}~\bibnamefont {Nakamura}},
  \bibinfo {author} {\bibfnamefont {K.}~\bibnamefont {Ohgushi}}, \ and\
  \bibinfo {author} {\bibfnamefont {T.}~\bibnamefont {Kimura}},\ }\href@noop {}
  {\bibfield  {journal} {\bibinfo  {journal} {Phys. Rev. B},\ }\textbf
  {\bibinfo {volume} {78}},\ \bibinfo {pages} {140401(R)} (\bibinfo {year}
  {2008})}\BibitemShut {NoStop}%
\bibitem [{\citenamefont {Mason}\ \emph {et~al.}(2006)\citenamefont {Mason},
  \citenamefont {Abernathy}, \citenamefont {Anderson}, \citenamefont {Ankner},
  \citenamefont {Egami}, \citenamefont {Ehlers}, \citenamefont {Ekkebus},
  \citenamefont {Granroth}, \citenamefont {Hagen}, \citenamefont {Herwig},
  \citenamefont {Hodges}, \citenamefont {Hoffmann}, \citenamefont {Horak},
  \citenamefont {Horton}, \citenamefont {Klose}, \citenamefont {Larese},
  \citenamefont {Mesecar}, \citenamefont {Myles}, \citenamefont {Neuefeind},
  \citenamefont {Ohl}, \citenamefont {Tulk}, \citenamefont {Wang},\ and\
  \citenamefont {Zhao}}]{Mason06}%
  \BibitemOpen
  \bibfield  {author} {\bibinfo {author} {\bibfnamefont {T.~E.}\ \bibnamefont
  {Mason}}, \bibinfo {author} {\bibfnamefont {D.}~\bibnamefont {Abernathy}},
  \bibinfo {author} {\bibfnamefont {I.}~\bibnamefont {Anderson}}, \bibinfo
  {author} {\bibfnamefont {J.}~\bibnamefont {Ankner}}, \bibinfo {author}
  {\bibfnamefont {T.}~\bibnamefont {Egami}}, \bibinfo {author} {\bibfnamefont
  {G.}~\bibnamefont {Ehlers}}, \bibinfo {author} {\bibfnamefont
  {A.}~\bibnamefont {Ekkebus}}, \bibinfo {author} {\bibfnamefont
  {G.}~\bibnamefont {Granroth}}, \bibinfo {author} {\bibfnamefont
  {M.}~\bibnamefont {Hagen}}, \bibinfo {author} {\bibfnamefont
  {K.}~\bibnamefont {Herwig}}, \bibinfo {author} {\bibfnamefont
  {J.}~\bibnamefont {Hodges}}, \bibinfo {author} {\bibfnamefont
  {C.}~\bibnamefont {Hoffmann}}, \bibinfo {author} {\bibfnamefont
  {C.}~\bibnamefont {Horak}}, \bibinfo {author} {\bibfnamefont
  {L.}~\bibnamefont {Horton}}, \bibinfo {author} {\bibfnamefont
  {F.}~\bibnamefont {Klose}}, \bibinfo {author} {\bibfnamefont
  {J.}~\bibnamefont {Larese}}, \bibinfo {author} {\bibfnamefont
  {A.}~\bibnamefont {Mesecar}}, \bibinfo {author} {\bibfnamefont
  {D.}~\bibnamefont {Myles}}, \bibinfo {author} {\bibfnamefont
  {J.}~\bibnamefont {Neuefeind}}, \bibinfo {author} {\bibfnamefont
  {M.}~\bibnamefont {Ohl}}, \bibinfo {author} {\bibfnamefont {C.}~\bibnamefont
  {Tulk}}, \bibinfo {author} {\bibfnamefont {X.~L.}\ \bibnamefont {Wang}}, \
  and\ \bibinfo {author} {\bibfnamefont {J.}~\bibnamefont {Zhao}},\ }\href@noop
  {} {\bibfield  {journal} {\bibinfo  {journal} {Physica B},\ }\textbf
  {\bibinfo {volume} {385-386}},\ \bibinfo {pages} {955} (\bibinfo {year}
  {2006})}\BibitemShut {NoStop}%
\bibitem [{\citenamefont {Copley}\ and\ \citenamefont {Cook}(2003)}]{Copley03}%
  \BibitemOpen
  \bibfield  {author} {\bibinfo {author} {\bibfnamefont {J.~R.~D.}\
  \bibnamefont {Copley}}\ and\ \bibinfo {author} {\bibfnamefont {J.~C.}\
  \bibnamefont {Cook}},\ }\href@noop {} {\bibfield  {journal} {\bibinfo
  {journal} {Chem. Phys.},\ }\textbf {\bibinfo {volume} {292}},\ \bibinfo
  {pages} {447} (\bibinfo {year} {2003})}\BibitemShut {NoStop}%
\bibitem [{\citenamefont {Azuah}\ \emph {et~al.}(209)\citenamefont {Azuah},
  \citenamefont {Kneller}, \citenamefont {Qiu}, \citenamefont
  {Tregenna-Piggott}, \citenamefont {Brown}, \citenamefont {Copley},\ and\
  \citenamefont {Dimeo}}]{Dave}%
  \BibitemOpen
  \bibfield  {author} {\bibinfo {author} {\bibfnamefont {R.~T.}\ \bibnamefont
  {Azuah}}, \bibinfo {author} {\bibfnamefont {L.~R.}\ \bibnamefont {Kneller}},
  \bibinfo {author} {\bibfnamefont {Y.}~\bibnamefont {Qiu}}, \bibinfo {author}
  {\bibfnamefont {P.~L.~W.}\ \bibnamefont {Tregenna-Piggott}}, \bibinfo
  {author} {\bibfnamefont {C.~M.}\ \bibnamefont {Brown}}, \bibinfo {author}
  {\bibfnamefont {J.~R.~D.}\ \bibnamefont {Copley}}, \ and\ \bibinfo {author}
  {\bibfnamefont {R.~M.}\ \bibnamefont {Dimeo}},\ }\href
  {http://nvl.nist.gov/pub/nistpubs/jres/114/6/V114.N06.A04.pdf} {\bibfield
  {journal} {\bibinfo  {journal} {J. Res. Natl. Inst. Stan. Technol.},\
  }\textbf {\bibinfo {volume} {114}},\ \bibinfo {pages} {341} (\bibinfo {year}
  {209})}\BibitemShut {NoStop}%
\bibitem [{\citenamefont {Fishman}\ \emph {et~al.}(2008)\citenamefont
  {Fishman}, \citenamefont {Ye}, \citenamefont {Fernandez-Baca}, \citenamefont
  {Haraldsen},\ and\ \citenamefont {Kimura}}]{Fish08}%
  \BibitemOpen
  \bibfield  {author} {\bibinfo {author} {\bibfnamefont {R.~S.}\ \bibnamefont
  {Fishman}}, \bibinfo {author} {\bibfnamefont {F.}~\bibnamefont {Ye}},
  \bibinfo {author} {\bibfnamefont {J.~A.}\ \bibnamefont {Fernandez-Baca}},
  \bibinfo {author} {\bibfnamefont {J.~T.}\ \bibnamefont {Haraldsen}}, \ and\
  \bibinfo {author} {\bibfnamefont {T.}~\bibnamefont {Kimura}},\ }\href@noop {}
  {\bibfield  {journal} {\bibinfo  {journal} {Phys. Rev. B},\ }\textbf
  {\bibinfo {volume} {78}},\ \bibinfo {pages} {140407(R)} (\bibinfo {year}
  {2008})}\BibitemShut {NoStop}%
\bibitem [{\citenamefont {Fishman}\ and\ \citenamefont
  {Okamoto}(2010)}]{Fish10}%
  \BibitemOpen
  \bibfield  {author} {\bibinfo {author} {\bibfnamefont {R.~S.}\ \bibnamefont
  {Fishman}}\ and\ \bibinfo {author} {\bibfnamefont {S.}~\bibnamefont
  {Okamoto}},\ }\href@noop {} {\bibfield  {journal} {\bibinfo  {journal} {Phys.
  Rev. B},\ }\textbf {\bibinfo {volume} {81}},\ \bibinfo {pages} {020402(R)}
  (\bibinfo {year} {2010})}\BibitemShut {NoStop}%
\bibitem [{\citenamefont {Haraldsen}\ and\ \citenamefont
  {Fishman}(2009)}]{Har09b}%
  \BibitemOpen
  \bibfield  {author} {\bibinfo {author} {\bibfnamefont {J.~T.}\ \bibnamefont
  {Haraldsen}}\ and\ \bibinfo {author} {\bibfnamefont {R.~S.}\ \bibnamefont
  {Fishman}},\ }\href@noop {} {\bibfield  {journal} {\bibinfo  {journal} {J.
  Phys.: Condens. Matter},\ }\textbf {\bibinfo {volume} {21}},\ \bibinfo
  {pages} {216001} (\bibinfo {year} {2009})}\BibitemShut {NoStop}%
\bibitem [{\citenamefont {Zhitomirsky}\ and\ \citenamefont
  {Zaliznyak}(1996)}]{Zhito96}%
  \BibitemOpen
  \bibfield  {author} {\bibinfo {author} {\bibfnamefont {M.~E.}\ \bibnamefont
  {Zhitomirsky}}\ and\ \bibinfo {author} {\bibfnamefont {I.~A.}\ \bibnamefont
  {Zaliznyak}},\ }\href@noop {} {\bibfield  {journal} {\bibinfo  {journal}
  {Phys. Rev. B},\ }\textbf {\bibinfo {volume} {53}},\ \bibinfo {pages} {3428}
  (\bibinfo {year} {1996})}\BibitemShut {NoStop}%
\bibitem [{\citenamefont {Haraldsen}\ \emph {et~al.}(2010)\citenamefont
  {Haraldsen}, \citenamefont {Ye}, \citenamefont {Fishman}, \citenamefont
  {Fernandez-Baca}, \citenamefont {Yamaguchi}, \citenamefont {Kimura},\ and\
  \citenamefont {Kimura}}]{Har10b}%
  \BibitemOpen
  \bibfield  {author} {\bibinfo {author} {\bibfnamefont {J.~T.}\ \bibnamefont
  {Haraldsen}}, \bibinfo {author} {\bibfnamefont {F.}~\bibnamefont {Ye}},
  \bibinfo {author} {\bibfnamefont {R.~S.}\ \bibnamefont {Fishman}}, \bibinfo
  {author} {\bibfnamefont {J.~A.}\ \bibnamefont {Fernandez-Baca}}, \bibinfo
  {author} {\bibfnamefont {Y.}~\bibnamefont {Yamaguchi}}, \bibinfo {author}
  {\bibfnamefont {K.}~\bibnamefont {Kimura}}, \ and\ \bibinfo {author}
  {\bibfnamefont {T.}~\bibnamefont {Kimura}},\ }\href@noop {} {\bibfield
  {journal} {\bibinfo  {journal} {Phys. Rev. B},\ }\textbf {\bibinfo {volume}
  {82}},\ \bibinfo {pages} {020404(R)} (\bibinfo {year} {2010})}\BibitemShut
  {NoStop}%
\bibitem [{\citenamefont {Plumer}(2007)}]{Plu07}%
  \BibitemOpen
  \bibfield  {author} {\bibinfo {author} {\bibfnamefont {M.~L.}\ \bibnamefont
  {Plumer}},\ }\href@noop {} {\bibfield  {journal} {\bibinfo  {journal} {Phys.
  Rev. B},\ }\textbf {\bibinfo {volume} {76}},\ \bibinfo {pages} {144411}
  (\bibinfo {year} {2007})}\BibitemShut {NoStop}%
\end{thebibliography}

%

\end{document}